\documentclass{interact}

\usepackage{natbib}
\bibpunct[, ]{(}{)}{;}{a}{}{,}

\usepackage{amsmath,amssymb}
\usepackage{graphicx}
\usepackage[normalem]{ulem}
\usepackage{hyperref}

\graphicspath{{./}{fig/}}

\newcommand{\sigv}[1]{\ifdim #1pt<0.05pt \underline{#1} \else%
                      \ifdim #1pt=0.05pt \underline{#1} \else%
                      #1 \fi \fi}
\newcommand{\sigi}[1]{\ifdim #1pt<0.01pt \underline{#1} \else%
                      \ifdim #1pt=0.01pt \underline{#1} \else%
                      #1 \fi \fi}
                      
\newcommand{\margv}[1]{\ifdim #1pt<0.05pt \underline{#1} \else%
                       \ifdim #1pt=0.05pt \underline{#1} \else%
                       \ifdim #1pt<0.10pt \uwave{#1} \else%
                       \ifdim #1pt=0.10pt \uwave{#1} \else%
                       #1 \fi \fi \fi \fi}
\newcommand{\margi}[1]{\ifdim #1pt<0.01pt \underline{#1} \else%
                       \ifdim #1pt=0.01pt \underline{#1} \else%
                       \ifdim #1pt<0.05pt \uwave{#1} \else%
                       \ifdim #1pt=0.05pt \uwave{#1} \else%
                       #1 \fi \fi \fi \fi}

\newcommand{\rpkg}[1]{{\em #1}}

\begin{document}

\title{Evaluating Spacing Tests of Multi-Modality}
\author{
  \name{Greg~Kreider}
  \thanks{CONTACT Greg Kreider.  Email: gkreider@primachvis.com}
  \affil{\small Primordial Machine Vision Systems, Inc.} }

\maketitle

\begin{abstract}
The spacing of data contains information about the underlying modality.
Sections of consistent, similar spacing correspond to modes while it
increases between them.  We have defined parametric, runs-based
non-parametric, and data driven tests to evaluate these features ---
flats and peaks --- and determine the presence and location of
multiple modes.  This report will check the tests in two situations.
By varying a bi-modal setup we can control the changes in spacing and
determine the resolution and sensitivity of the tests.  By applying
them to the large number of test cases that exist in the literature
from other modality studies we check the stability and consistency
of the results. We will also evaluate the accuracy of the
null-distribution models of the features.  The results show that the
spacing does reflect the data's modality, that the tests do screen
marginal cases, and where the analysis begins to break down.
\end{abstract}

\begin{amscode}
62E10 (primary), 62G30 (secondary)
\end{amscode}

\section{Introduction}
\label{sec:intro}

Spacing, the difference between data points after sorting, has
occasionally been used to analyze the data's modality.
\citep{venter67} locates the mode at the minimum of what we will
call here the interval spacing between non-adjacent order
statistics, with the interval width reflecting the distribution's
curvature at that point.  \citep{rufibach10}, extending
\citep{duembgen08}, uses an ensemble of interval spacings to test
for regions of non-zero slope.  These two examples use one
feature of the spacing, that at modes it has similar values.  It
appears consistent or stable.  If small the data is dense.  Away
from the mode the spacing increases strongly in the tails of the
distribution.  Between two modes this becomes a local rise.  The
appearance of either feature, flat sequences or local peaks, not
only signals that the data has a multi-modal distribution, but
also locates where the distributions change.  However, although
we know an expression for the expected spacing for any ensemble
of variates, in practice it can only be solved numerically and
we have no theoretical description of the features, even for as
simple a setup as two normal draws.  Determining when they
signal multi-modality, which setups they can resolve, and how
best to quantify their occurrence is the challenge.

We have defined a variety of tests to evaluate these peaks and
flats.  They include parametric models based on a univariate
null distribution, using the count or length of signed runs
within the interval spacing, feature reconstruction by a
permutation of those runs or by bootstrap sampling from the
differential spacing, and a fusion of existing changepoint
detectors to identify where the behavior of the spacing
changes.

Section~\ref{sec:di} describes the features at a high level and
Section~\ref{sec:tests} the tests.

These tests have different strengths and weaknesses, and the goal
of this paper is to evaluate their performance.
Section~\ref{sec:models} compares the feature models to the data
they fit, checking their accuracy, the bounds in their parameter
space beyond which they deteriorate, and misclassification rates.
Section~\ref{sec:bimodal} systematically varies the parameters of
two normal variates in order to investigate the detection of the
features, the accuracy of their placement, and the significance
level of the various tests.  We will see that flats mark modes
and peaks anti-modes, albeit with decreasing stability as it
becomes harder to distinguish the two variates.  Changepoints
fall in-between, responding not to the features themselves but
to the transition between them.  The models and bootstrap tests
on the spacing after low-pass filtering give a clear signal when
the data is multi-modal.  The runs tests on the interval spacing
can only find much larger differences and respond slowly to
changes.  The section finishes by comparing the spacing analysis
to other modality checks, finding that the low-pass spacing is a
little more sensitive and the interval spacing a little less,
with the runs tests also being less decisive.  The literature
contains a large number of test samples to evaluate these
existing checks, and Section~\ref{sec:litds} runs the spacing
analysis on repeated draws of each, looking at the stability of
the features, the accuracy of their placement, and comparing
low-pass to interval features.  Although the wide variety of
setups prevents general conclusions, some results are notable.
Small samples next to larger vary between draws and reduce the
stability of peaks.  Mode transitions at the edge of the sample
are lost during low-pass filtering but can appear in the
interval spacing.  The roughness of that signal means peaks are
less stable and repeatable and do not match well to the low-pass
features, especially after testing.  They do not align as
strongly to the anti-modes.  Flats in both spacings do agree,
however.

This paper is written as a companion to \citep{kreider25a}, which
fully defines the feature detectors and each test but only
summarizes the evaluation done here.  \citep{kreider25c} discusses
the reference implementation available in the R package
\rpkg{Dimodal} \citep{kreider25e} and as a practical example applies
the analysis to identify Kirkwood gaps in the orbits of asteroids.

\section{Spacing Features and Modality}
\label{sec:di}

The spacing of a variate takes a distinctive form, a `U' with a broad
base at the center and sharply increasing sides from the tails
\citep[Figure 1]{kreider23a}.  Using the notation from Pyke \citep{pyke65},
the spacing is the difference between consecutive order statistics
$ T_{i} $, or $ D_{i} = T_{i} - T_{i-1} $ for upper index
$ i = 2 \ldots n $ of a sample with size $ n $.  Its density
\begin{equation} \label{eq:fdi}
f_{D_i}(y) = \frac{n!}{(i-2)! ~ (n-i)!}
  \int_{-\infty}^{\infty}
    \left\{ F(x) \right\}^{i-2} \left\{ 1 - F(x+y) \right\}^{n-i} ~
    f(x) f(x+y) ~ dx
\end{equation}    
follows from the order statistic density for a variate with distribution
or cumulative density function $ F(x) $ and density $ f(x) $.  The
expected spacing, solved normally by integrating the first moment for
non-negative $ y $, has a closed form for only a few variates:
uniform, exponential and logistic.
\begin{align} \label{eq:edi}
E\Bigl\{ D_{i} \Bigr\} & = \int_{0}^{\infty} y ~ f_{D_{i}}(y) ~ dy \\
E\Bigl\{ D_{i,unif} \Bigr\} & = \frac{b - a}{n + 1} \nonumber \\
E\Bigl\{ D_{i,exp} \Bigr\} & = \frac{1}{\lambda (n-i+1)} \nonumber \\
E\Bigl\{ D_{i,logis} \Bigr\} & = \frac{\sigma n}{(i-1) (n-i+1)} \nonumber
\end{align}
There is also a series expansion for Gumbel variates.  The uniform and
exponential results are well-known and can be found in \citep{pyke65},
the logistic and Gumbel equations in \citep{kreider23a}.  The flat
bottom of the `U' is seen by solving these equations for the range of
indices where the expected spacing remains within a factor $ \alpha $
of the minimum, which is $ (n \alpha + 1) / (\alpha + 1) $ for
exponential variates and $ 1 + n \sqrt{\alpha / (\alpha + 1)} $ for
logistic.  The scaling at the first and last points is $ n-1 $ and
$ n^{2}/4(n-1) $ times larger than the minimum, respectively.  For a
draw of $ n = 100 $ points, for example, the spacing of an
exponential variate stays within 10\% of the minimum for 10 points
and increases by a factor of 99 in the tails.  For a logistic sample
the spacing is stable over 31 points and increases $ 25 \times $.
The `U' indeed has steep sides and a wide bottom. The spacing's
variance is high.  For an exponential variate it is the square of
the expected value.  This also seems to be true of logistic samples,
and although there is an analytic expression involving several
series expansions, it has not been simplified to demonstrate the
equivalence.

\begin{figure}
\centering
\begin{minipage}[t]{\textwidth}
\includegraphics[width=\textwidth]{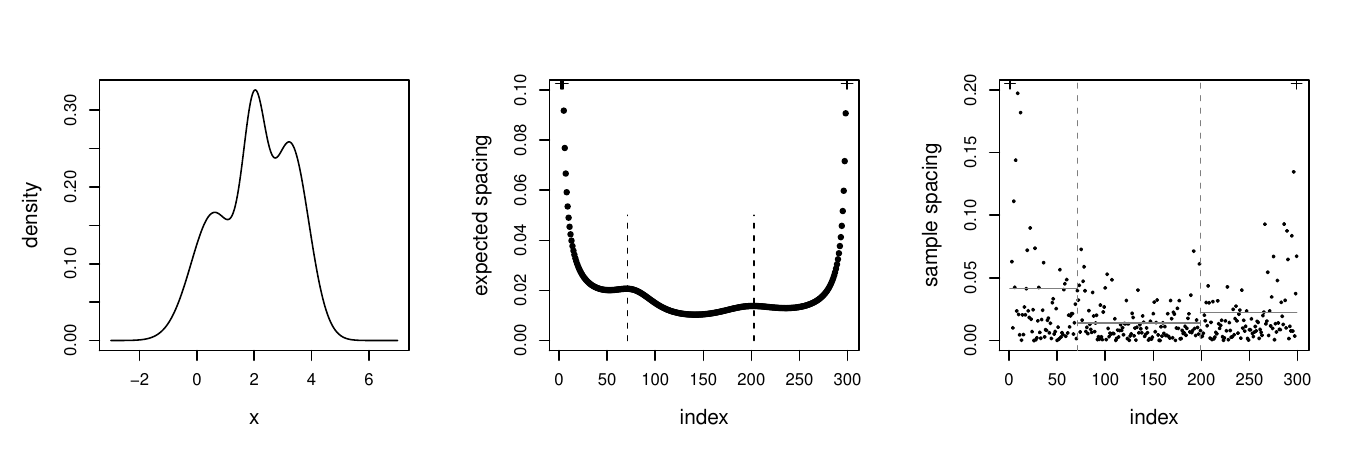}
\caption{\label{fig:diex} Tri-modal density function (left), expected
 spacing (middle), and sample draw (right). }
\end{minipage}
\end{figure}

The spacing of multi-modal setups also follows \eqref{eq:fdi}.  The
distribution and density functions are the sum of the individual
variates weighted by the draw size.  Closed form solutions for the
expected spacing are not possible, and modeling the combined spacing
breaks down where the modes interact.  Numeric solutions are of course
possible but deliver no insight into the behavior.  Qualitatively,
however, we can describe what happens.  Within the modes the spacing
will be stable at levels that depend on distributional parameters,
like the exponential rate $ \lambda $ or logistic scale $ \sigma $
in \eqref{eq:edi}.  Between modes the spacing will increase, with
the size and location of the local maximum depending on the location
and scale of the variates.  Figure~\ref{fig:diex} provides an
example.  The density in the left graph is a combination of three
normal draws,
\begin{equation} \label{eq:triex}
 100 \times N(0.6,0.8), \quad 75 \times N(2.0,0.4),
 \quad 125 \times N(3.25,0.65)
\end{equation}
where the second parameter is the standard deviation.  The expected
spacing in the middle graph shows two small increases at indices
70 and 202, corresponding to $ x = 1.01 $ and 2.78.  The positions
differ slightly from the minima in the density at 1.06 and 2.74,
caused by the difference in standard deviation and, to a lesser
extent, the draw size. We will see this shift in
Section~\ref{sec:bimodal}.  These same forces pull the peaks away
from the midpoint of the means, particularly for the first pair.
Capturing such differences would be the goal of a theoretical study
of the spacing; it is what we lose by not being able to solve the
expected spacing in general.  Note that the $ y $ axis is restricted
to show the features and not the initial or final tails, with
crosses marking points that fall outside the graphs.  The spacings
at each side are 0.284 and 0.228, more than ten times larger
than the minimum.  The right graph plots the spacing for a sample
draw from \eqref{eq:triex}.  The average values within each region,
of 0.041, 0.014, and 0.023, roughly reflect the standard deviations.
The increase at index 70 is caused by a lack of small spacing values
and that at 202 by locally larger spacings.

By taking the difference beyond adjacent order statistics we get the
``interval spacing''.  With an extension of the notation,
$ D_{i,w} = T_{i} - T_{i-w} $, where $ w $ is the width of the interval
and $ i $ the upper index ranges from $ w + 1 $ to $ n $.  The interval
spacing is the sum of individual spacings,
\begin{align} \label{eq:diw}
\sum_{j=0}^{w-1} D_{i-j}
  = ~ & (T_{i} - T_{i-1}) + (T_{i-1} - T_{i-2}) + \ldots
      + (T_{i-w+1} - T_{i-w}) \nonumber \\
  = ~ & T_{i} - T_{i-w} = D_{i,w}
\end{align}
But this sum is the same as applying a rectangular or running mean
filter to the spacing.  The rectangular filter is simple and its large
main lobe allows the use of smaller widths than other low-pass filters.
It has relatively large side lobes, however, which do not suppress 
high-frequency components well, so the interval spacing will be rough.
Not normalizing \eqref{eq:diw} by the width appears to amplify the
differences and makes this unevenness worse.  The density of the
interval spacing is known, for example derived
from \citep[(2.4)]{pyke65}
\begin{align} \label{eq:fdiw}
f_{D_{i,w}}(y) & =
\begin{aligned}[t]
  S_{1} \int_{-\infty}^{\infty} &
  \left\{ F(x) \right\}^{i-w-1} \left\{ F(x+y) - F(x) \right\}^{w-1}
  \left\{ 1 - F(x+y) \right\}^{n-i} \\
  & \times f(x) f(x+y) ~ dx
\end{aligned} \\
S_{1} & = \frac{n!}{(i-w-1)! ~ (w-1)! ~ (n-i)!} \nonumber
\end{align}
Although a closed-form solution for uniform, exponential, and logistic
variates is possible for this and its moments, the equations involve
series expansions and are not particularly helpful.  Evaluating the
series must be done with high-precision math libraries because their
sum must be close to zero to counter-balance the scaling of $ S_{1} $,
especially near the middle of the sample.  Once again studying
multi-modal setups is not possible without numeric integration.

Better smoothing of the spacing is possible with other low-pass filters.
The Finite Impulse Response family is a good choice, for there are no
uncompensated poles to affect stability.  \citep{harris78} has a
comprehensive overview of the various FIR kernels and concludes that
the Kaiser filter \citep[Section 7.4.3]{oppenheim89} is best in general.
This matches our experience during development, although in practice
all perform similarly. Unlike the upper indexing of the interval
spacing, we center the filter on each data point, which aligns
features with the underlying data; the interval spacing must be shifted
by $ w / 2 $ for matching.  We choose to not filter at indices where the
kernel extends outside the data, that is, within half the kernel size of
the first or last point, because there is no good way to estimate the
continued growth of the spacing further into the tails.

The peak detector begins by identifying local minima and maxima in a
signal, the points that are larger or smaller than both neighbors.
It calculates the height or value difference of the maxima to the
surrounding minima, keeping the smaller of the two.  Working in
increasing height order, it removes those maxima that are subsidiary
to larger peaks, or equivalently that have shallow minima separating
them.  This happens when the height is less than a fraction
$ f_{ht} $ of the range of the data or if its relative height,
compared to the average value of the peak and minima, is less than
a fraction $ f_{relht} $.  In other words, remove a peak and minima
with values $ x_{pk} $ and
$ x_{i} $ if $ x_{pk} - x_{i} \le f_{ht} ~ (\max(x) - \min(x)) $
or if $ \lvert x_{pk} - x_{i} \rvert /
          ((\lvert x_{pk} \rvert + \lvert x_{i} \rvert ) / 2)
       \le f_{relht} $.
The deletion changes the height of the neighboring maximum, requiring
the re-calculation of its height.  The detector keeps the initial and
final data point as the appropriate extremum, never merging them, and
also places the peak in the middle of tied or nearly equal sequences
of data to avoid noise if the peak has a flat top, such as a square
wave.  The final height of the remaining peaks taken to the smaller
of their bordering minima and scaled by the signal's standard
deviation is the characteristic value of the feature.

Flats use a traditional ripple specification to fix their height as a
fraction $ f_{ripple} $ of the data range.  For each data point the
detector determines the bounds of the ripple and scans outward in both
directions until the signal exceeds them.  To improve robustness the
detector can allow a few outliers.  The detector then works from the
longest flat to shortest.  It counts the number of points within the
flat that have not yet been assigned to one, and reports it if that
count is above a threshold, either an absolute length $ L_{min} $
or a relative fraction $ f_{minlen} $ of the amount of data.  Flats
can therefore overlap as long as the uncovered portion is longer
than the cut-off.  Because the ripple fixes the height of the flat,
its length becomes the characteristic value for judging the feature.

A more general approach looks for changes in the behavior of data.
Called changepoint detection, the technique has a long, rich
history, with roots in statistical process control and economic
trend analysis.  More than 25 maintained packages are available in
R.  They use a variety of approaches, including two-sample and
cumulative sum tests, partitioning schemes based on penalty
functions usually involving information criteria, outlier detection
using parametric models of the data, and regression fits.
Changepoints will not identify the modes and anti-modes directly,
since they respond to the transitions between them.  If the change
is abrupt, for example in some literature test cases where a very
tight variate is added to a broad background, the detectors can
locate the edge of feature. Real data will not usually see such
boundaries.  We also cannot evaluate the quality of the final
changepoint list, since the libraries rarely provide this
information and any test we could add would duplicate one used by
some algorithm.  The lack of this quantitative evaluation limits
our ability to fuse the individual classifiers, leaving majority
voting as the only possibility \citep{ruta00} \citep{xu92}.
Changepoints supplement the peaks and flats, and then somewhat
indirectly, since they will not match up.

In summary, studying modality using spacing involves first reducing
the high variability, either by low-pass filtering or taking intervals
with large widths, and then applying the feature detectors.  Peaks
will indicate multi-modality and will align with anti-modes, while
flats will cover the modes.  A combination of changepoint detectors
run on the raw spacing will complement these features.

\section{Tests of Features}
\label{sec:tests}

Knowing the distribution of the spacing might allow us to model the
features, for example using Brownian excursions were the spacing to be
normally distributed (which it is not), and develop tests of their
significance.  Since this is not possible our approach evolved to
minimize assumptions.  The first tests model the distribution of peak
heights and flat lengths from a null variate, similar to existing
modality tests based on the normal or uniform distribution.  The
second set applies non-parametric tests to runs of increasing or
decreasing or tied spacing, using a combinatorial analysis for the
number of runs or a Markov chain model for the length of the longest.
The third set uses data-driven tests, either permutations of the
runs to re-create a peak or a bootstrap from the difference of the
smoothed spacing to build either a peak or a flat.  Finally, the
fusion of the changepoint detectors requires a voting algorithm that
can handle small uncertainties in the point locations.

The changepoint detector combines the results of whatever libraries
are available on the system.  A library may contain multiple
algorithms or tests, and the detector combines these into a set of
library changepoints. It merges these into a master list,
considering points near each other to be the same.  As a consistency
requirement it ignores libraries that identify unusually few or
many points, either as a fraction of the data, more than 5\%, or
relative to the other libraries, typically outside 10--90\% of all
counts.  The detector reports those points in the master set that
appear in a majority of the remaining libraries.

The parametric models are built from the features found in
uni-variate samples, gathering the heights and lengths generated.
They take as model parameters the size of the draw $ n $, the
low-pass filter type, its kernel size $ f_{lp} $ as a fraction of
$ n $, quantiles $ q $ ranging from 0.9 to 0.99999 (which we will
denote as 0.9x5 to avoid counting nines), and the null or base
distribution.  The quantile range requires a million samples be
drawn for stability in the critical values.  The models are
created for the best fit and have no theoretical basis.

For a given filter size and quantile the standardized peak height is
proportional to $ \log n $, but there is no pattern to the interaction
of all three parameters.  We model the distribution of heights and
find that an inverse Gaussian or Wald gives the best fit.  Its
location parameter $ \mu $ and shape parameter $ \lambda $ depend on
$ \log n $ and polynomially on $ f_{lp} $, with the test statistic,
a modified height, polynomial in $ f_{lp} $.  An inversion of the
model using the Wald quantile function gives critical values.  Note
that the inverse Gaussian also describes a Brownian excursion,
although a peak is not an excursion since it is not required to
return to its baseline. The preferred low-pass filter, the Kaiser,
is a good conservative choice, producing fewer and smaller peaks
than other kernels. Their critical values are roughly constant
multiples of the Kaiser and separate models are not needed.  There
is also a conservative choice of null distribution, an asymmetric
Weibull variate with scale parameter $ a = 2 $ and shape $ b = 4 $.
Its peak height at $ q = 0.95 $ appears in other variates at the
0.975 or 0.99 level, except for two outlier distributions, the beta
and uniform.  These produce so many and so large peaks that their
0.95 quantiles match the other distributions at the 0.9x5 level.

In some ways the flat model is the opposite of the peak.  It is simpler
and fits better, but the choice of null distribution is not as clear.
The model contains 24 terms with interactions linear in $ f_{lp} $,
quadratic in $ n $, and quadratic in $ q $ plus a logistic factor.
Regression determines the coefficients of the terms.  No mapping through
a distribution function is needed, and critical values are found
numerically with a root solver.  The Kaiser filter is still the most
conservative choice, but there is a larger variation for other filters
and they cannot be handled by scaling.  Each needs a set of model
coefficients.  The flat lengths do depend on the null distribution,
and there is not much overlap between them. Three distributions, the
Weibull, logistic, and Gumbel, span the range of critical values
from shortest to longest.  Separate models must be built for each.
The logistic produces quantiles in-between the others and is the
default choice.

If a sequence of values is built from a number of discrete symbols,
in our case by taking the sign of the difference of consecutive
interval spacings giving $ -1 $, $ +1 $, or 0 if there are ties,
then \citep{kaplansky45} showed by a combinatorial counting that
the number of runs of consecutive symbols is distributed normally
in the asymptotic limit, and provides expressions for the expected
count and its variance that depend on the symbol populations.
Simpler expressions apply if there are only two symbols
\citep{wald40}, which would happen if the data is continuous and
ties do not happen.  The Kaplansky-Riordan test compares the
number of runs within a peak against the expected.

The interval spacing introduces correlations into the signed
difference through the common span of consecutive intervals, which
a combinatorial counting ignores.  We can capture them in a Markov
chain model.  After splitting the inter-symbol transition matrix
into its diagonal, which advances a run within a symbol, and
off-diagonal elements, which ends a run, we can set up a second
Markov chain for the growth in length over all symbols.  The new
run transition matrix has the off-diagonal transitions in its
first column, the advancing sub-matrices next to the diagonal,
and an identity sub-matrix added as an absorbing state beyond the
longest length $ L $.  Evolving this chain over the size of the
feature then gives the probability of producing a run no longer
than that length.  The differences between run lengths $ L-1 $
and $ L $ removes the inequality, giving the probability of the
longest run \citep{kreider25d}.  The calculation involves a
recursion of the split sub-matrices over the width of the peak.

A third test that can be made with the runs within a peak is to
permute them, re-construct the signal, and measure the resulting
height.  If the permutation separates the longer rising and
falling steps to each side of the maximum, with short opposing
runs in-between, it will form a high peak.  Mixing them produces
a smaller feature.  The height's quantile follows from the
distribution of the permutations, found by sampling unless the
number of runs is small enough for an exhaustive check.  The only
wrinkle to the test is an additional requirement that the
permutation cannot place runs of the same symbol side by side,
which would form longer runs than actually exist.

This approach can be taken in general to find the probability of
either a peak or flat given its height and length, in either the
low-pass or interval spacing.  This bootstrap or excursion test
draws from a pool of the difference between filtered points, the
derivative of the low-pass or interval spacing.  Because the
initial and final tails in the spacing are much larger than the
others, the test drops points on either side if they are the
largest spacings in the data.  This is equivalent to removing
outliers beyond three or four standard deviations, but only
from the tails.  Bootstrap tests sample with replacement, and
the draw size matches the feature's.  For flats this is its
length.  For peaks it taken over some support range to some
fraction of the height, similar to Full Width at Half Maximum.
A smaller support will better reflect the peak's size if a
minimum falls in the middle of a long flat.  The draws are made
several thousand times to get a stable distribution of the
heights, and the quantile follows from its height against this
distribution.

The appendices of \citep{kreider25a} contain pseudo-code for the
detector and test algorithms.

\section{Model Accuracy}
\label{sec:models}

\begin{figure}
\centering
\begin{minipage}[t]{\textwidth}
\includegraphics[width=\textwidth]{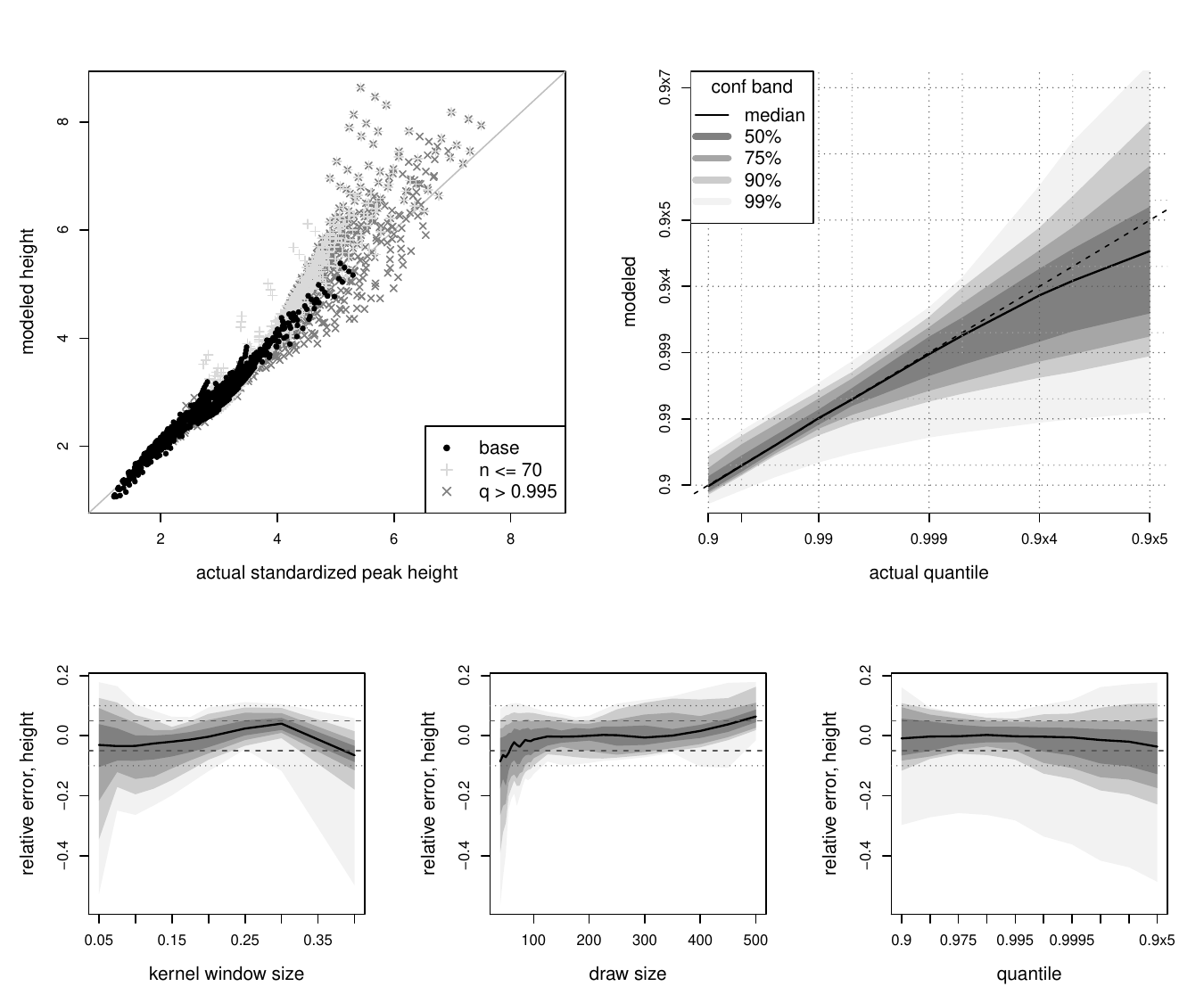}
\caption{\label{fig:htmodel} Peak height model accuracy and relative error
over model parameter space.
}
\end{minipage}
\end{figure}

The top graphs of Figure~\ref{fig:htmodel} compare the peak height
model to the training data in both directions, as quantile or
critical value.  The predicted value matches the actual for
standardized heights up to 4, for all model parameter combinations.
Beyond this it splits into three groups.  The central parameter
space remains accurate for all heights. When $ n \leq 70 $ the
models predict too large a height.  For $ q > 0.995 $ the
prediction still follows the ideal line but the envelope widens
and there can be large errors in the critical value.  This
uncertainty is less important as the peaks will still be accepted
at the usual 0.05 or 0.01 level.  The increased spread is clear
in the top right graph, where the confidence bands are taken over
the combinations of draw and filter size for each measured
quantile.  Beyond $ q = 0.999 $ the bands widen and the model
lags the actual a little.  Peaks appear less likely than they are
and the model is conservative.  The confidence bands translate
into misclassification rates.  The area above the modeled value
is the false positive rate, below the false negative.  For
example, the horizontal dotted line at the 0.95 modeled quantile
lies just beyond the 75\% confidence band of the 0.90 actual
quantile, which means that 12\% of the peaks, half the two-sided
range, will seem to be significant when they are not.
Table~\ref{tbl:htmiss} counts the wrong predictions at the 0.05
and 0.01 levels.  The 0.05 level is biased towards accepting
peaks, generating false positives, while the 0.01 level is
better balanced.

\begin{table}
\centering
\caption{\label{tbl:htmiss} Peak Height Missed Classification Rates}
{\small
\begin{tabular}{lrcccrccc}
actual $ q $ && \multicolumn{3}{c}{at $ q = 0.95 $} &&
  \multicolumn{3}{c}{at $ q = 0.99 $} \\
0.90    && FP & 12.0\% & (26/216) && FP &  0.0\% & \\
0.95    &&    &        &          && FP &  0.0\% & \\
0.975   && FN &  1.4\% & (3/216)  && FP &  8.8\% & (19/216) \\
0.99    && FN &  0.5\% & (1/216)  &&    &        & \\
0.995   && FN &  0.0\% &          && FN &  9.3\% & (20/216) \\
0.9x4   && FN &  0.0\% &          && FN &  1.4\% & (3/216)  \\
0.9x4+  && FN &  0.0\% &          && FN &  3.7\% & (8/216)  \\
\end{tabular}
}
\end{table}

The bottom row of Figure~\ref{fig:htmodel} plots the residuals
expressed as an error relative to the actual value,
$ \delta_{rel} = (h_{act} - h_{pred}) / h_{act} $.  Positive
values mean the model under-estimates the height.  The dashed
horizontal lines bound $ \pm 5\% $ differences, the dotted
$ \pm 10\% $.  Confidence bands are taken over the other two
model parameters.  The median predictions have a 5\% or smaller
relative error, and in general the predictions for 75\% of the
parameter combinations remain within 10\%.  The bands widen for
small kernels and draw sizes, or for larger quantiles.  

In other words, the height model is usable over the parameter
combinations.  It becomes increasingly inaccurate for small
data sets with fewer than 100 points or for kernel sizes less
than 10\% or more than 30\%.  The width of the confidence bands
and corresponding misclassification rates suggest the test
should be evaluated at the 0.01 level.

\begin{figure}[ht]
\centering
\begin{minipage}[t]{\textwidth}
\includegraphics[width=\textwidth]{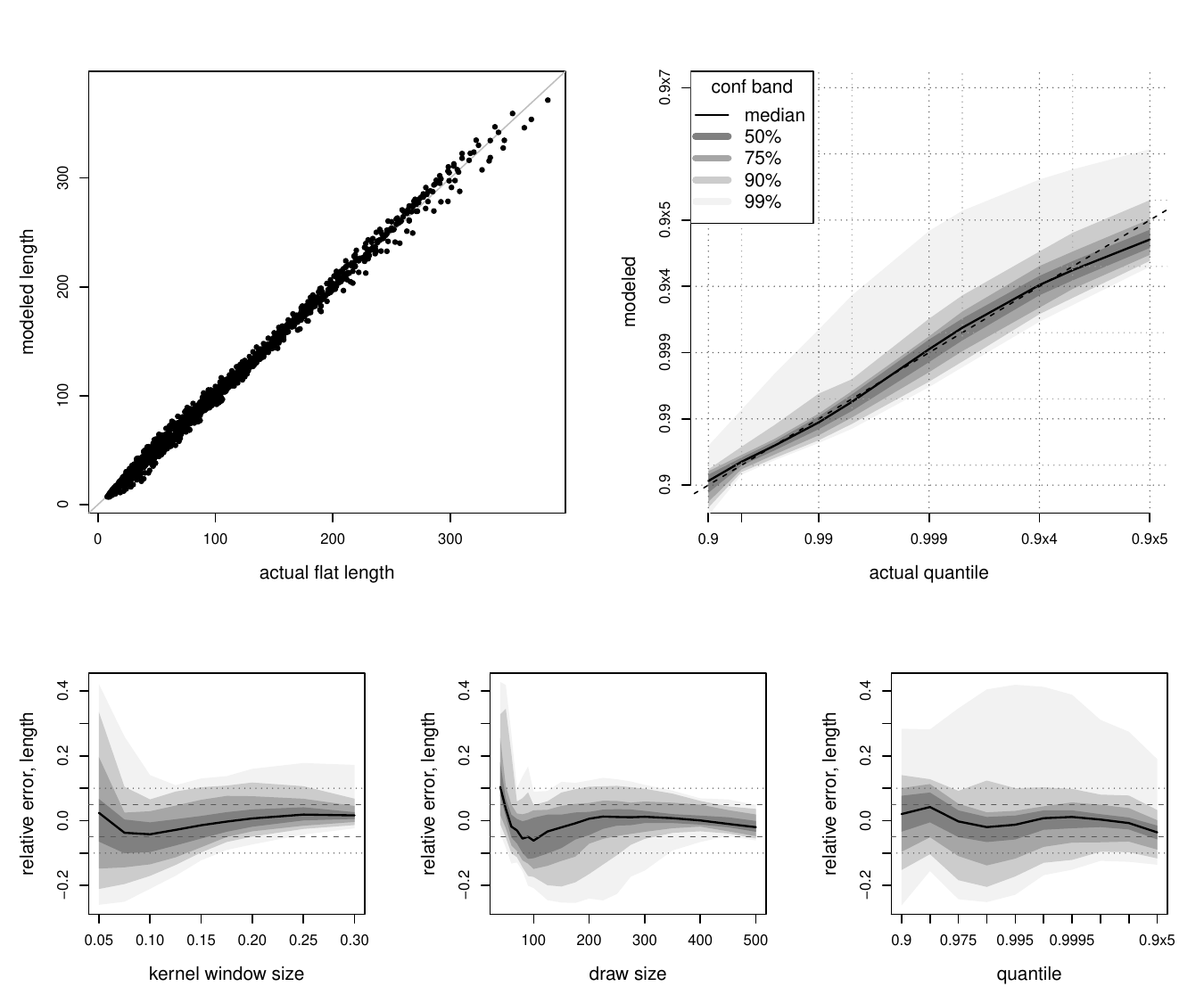}
\caption{\label{fig:lenmodel} Flat length model accuracy and
relative error over model parameter space.
}
\end{minipage}
\end{figure}

The accuracy of the flat length model is cleaner than for the peak
height (Figure~\ref{fig:lenmodel}).  The predicted length falls
along the ideal matching line for all combinations of model
parameters in the upper left graph, as does the predicted quantile
in the upper right.  There is no broadening of the confidence
range at larger quantiles.  The outliers forming the 99\%
confidence band occur for the smallest draw and kernel size,
$ n \leq 60 $ and $ f_{lp} \leq 0.075 $.  This means that the
misclassification rates in Table~\ref{tbl:lenmiss} are smaller.
The test can be evaluated at the 0.05 level.

\begin{table}[ht]
\centering
\caption{\label{tbl:lenmiss} Flat Length Missed Classification Rates}
{\small
\begin{tabular}{lrcccrccc}
actual $ q $ && \multicolumn{3}{c}{at $ q = 0.95 $} &&
  \multicolumn{3}{c}{at $ q = 0.99 $} \\
0.90    && FP &  1.3\% & (1/80)   && FP &  0.0\% & \\
0.95    &&    &        &          && FP &  1.3\% & (1/80) \\
0.975   && FN &  0.0\% &          && FP &  1.3\% & (1/80) \\
0.99    && FN &  0.0\% &          &&    &        & \\
0.995   && FN &  0.0\% &          && FN & 10.0\% & (8/80) \\
0.9x4   && FN &  0.0\% &          && FN &  0.0\% & \\
0.9x4+  && FN &  0.0\% &          && FN &  0.0\% & \\
\end{tabular}
}
\end{table}

The median relative error again stays within 5\% of the actual
flat length.  The form of the curves speaks to a higher order
residual that might be corrected.  These trends drive the 75\%
confidence band below the line for the 10\% relative error for
moderate draw sizes between 70 and 200 points, kernel sizes
below 0.15, and quantiles around 0.99.  At the smallest draws
and kernels the error grows rapidly, implying the length model
is breaking down in this region, but these conditions will
naturally create short flats that would be removed by the
minimum length requirement.  The errors remain small at the
upper edges of the model parameter space.

\section{Bi-Modal Variations}
\label{sec:bimodal}

\begin{figure}
\centering
\begin{minipage}[t]{\textwidth}
\includegraphics[width=\textwidth]{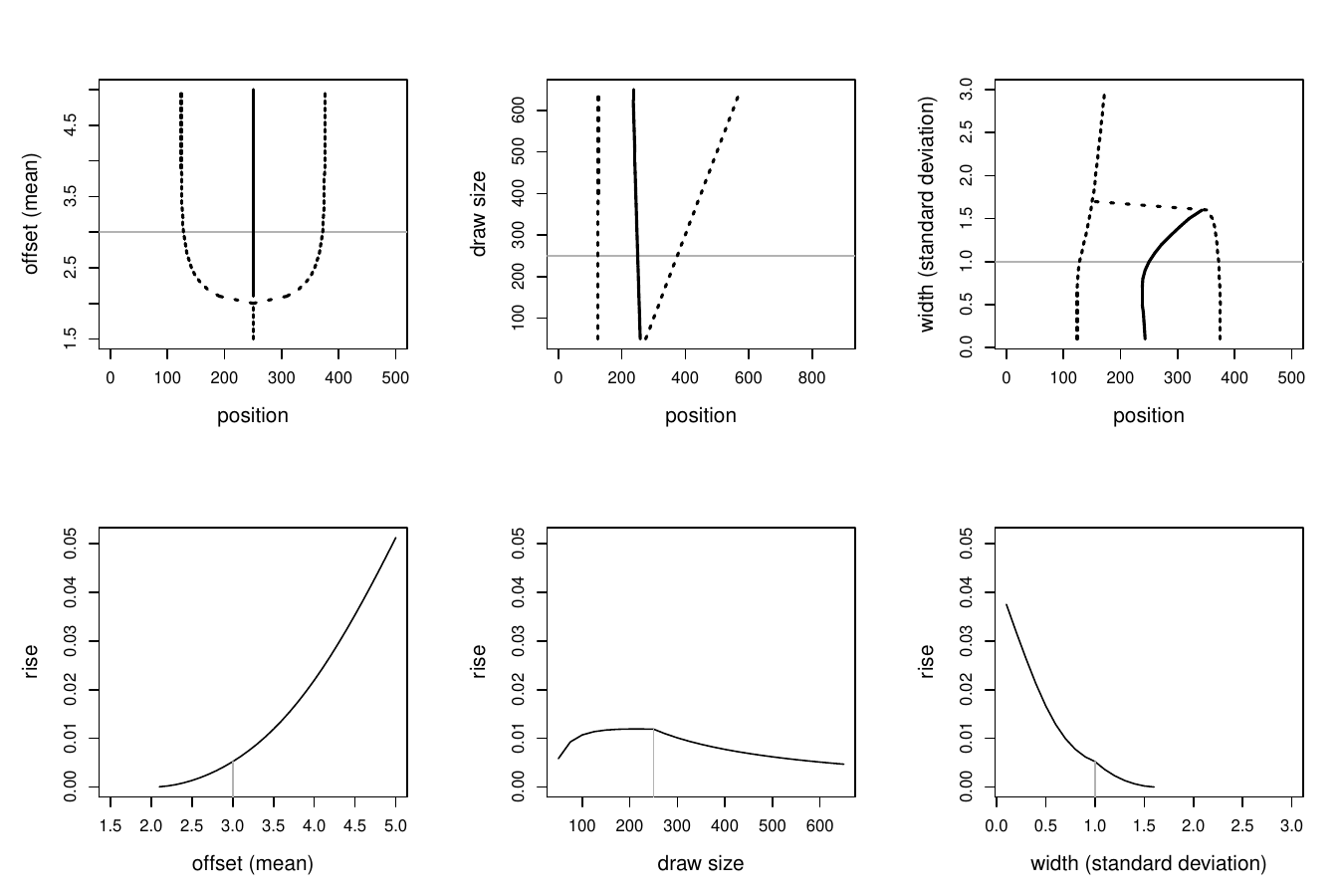}
\caption{\label{fig:bimodal} (Top) Position of modes (dotted lines) and
anti-mode (solid) for bi-modal variations.  (Bottom) Corresponding rise
in expected spacing at the anti-mode.
}
\end{minipage}
\end{figure}

To investigate how the tests behave under controlled conditions,
we vary a simple bi-modal distribution with two normal variates,
\begin{equation}
250 \times N(0,1), \quad n \times N(\mu, \sigma)
\end{equation}
$ \mu $ represents the separation between the variates and
$ \sigma $ the width of the second draw.  While changing one of the
parameters, the others will be held at $ n = 250 $ and
$ \sigma = 1.0 $.  The default separation for the draw variants is
3.3, for the others 3.0.  The larger value reflects a balance
between the ability of the tests to resolve the two draws, as we
will see.  We vary $ n $ between 50 and 650 in steps of 25, $ \mu $
between 1.5 and 5.0 in steps of 0.1, and $ \sigma $ between 0.1 and
3.0 in steps of 0.1. Solving \eqref{eq:edi} numerically for each
variant gives us the location of the modes and anti-mode separating
them, and also the local increase in the expected spacing.  We will
call this the rise to distinguish it from the peak height.  Because
the numeric integration is smooth we can find the extrema where the
differential signal changes sign, which will work in marginal cases
where the minimum height requirement of the peak detector would
fail.  The top row in Figure~\ref{fig:bimodal} plots the location
of the modes as dotted lines and the anti-mode as a solid. At some
point, for small separations or large standard deviations, it is
no longer possible to resolve the two variates.  The anti-mode
disappears at an offset of 2.0 and the modes merge into one.  The
setyp remains symmetrical in these variants and the anti-mode
stays at the central index.  More points in the second draw mostly
affect its mode, which tracks the middle of the draw and shifts
linearly to the right while slightly pushing the anti-mode into
the first variate.  A larger standard deviation turns the second
mode into a shoulder, until it can no longer be distinguished at
1.7.  The first mode's position drifts to the right as the second
weakens.  The flat for the second mode narrows rapidly beyond the
default variant, bounded by the anti-mode, although its position
is stable.  The bottom row of Figure~\ref{fig:bimodal} plots the
rise for each variant. At the default setup it is 0.0052, or
0.0089 for the size variants. It rises sharply as the offset
increases and width decreases.  An imbalance in the draw size
only decreases the rise, and rather slowly.  The balanced draw
produces the largest rise and is easiest to resolve.  Were we to
have used another separation the curve would shift vertically
without changing its shape.

\begin{figure}
\centering
\begin{minipage}[t]{\textwidth}
\includegraphics[width=\textwidth]{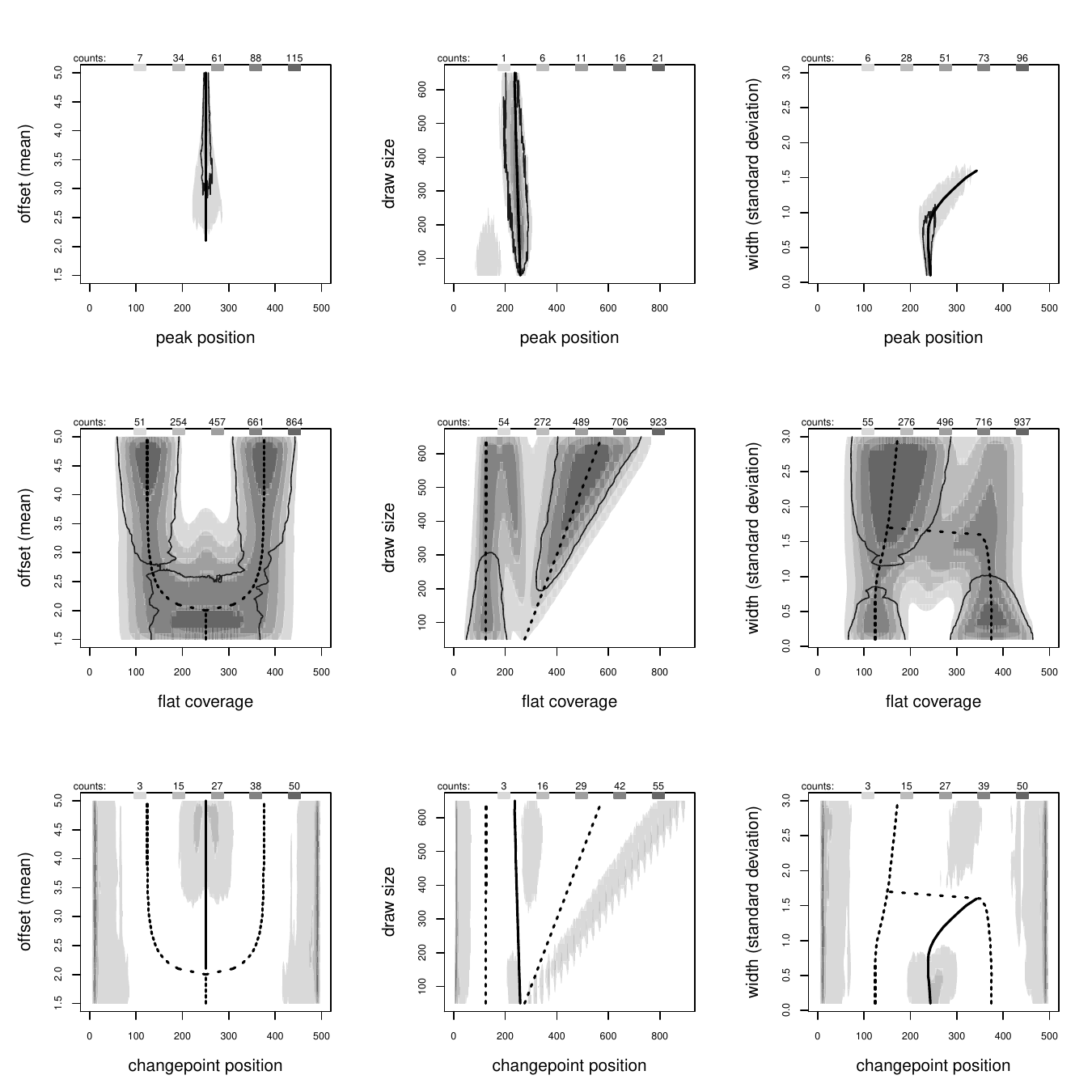}
\caption{\label{fig:lpfeat} Peak (top) and flat (middle) positions for
bi-modal variations after low-pass filtering, or changepoints (bottom)
in raw spacing.  Greyscale levels encode frequency of position over
1000 trials and differ between graphs.  Contour lines enclose 90\% of
accepted low-pass features.
}
\end{minipage}
\end{figure}

The bi-modal setup lets us study the response and stability of
features in the spacing.  We draw each variant one thousand times,
run the feature detectors on both smoothed spacings, and collect
their position and probability from each test.  We gather
changepoints for only half the trials because of the long run
time.  We use a Kaiser filter with kernel size $ f_{lp} = 0.15 $
of the total draw for the low-pass spacing and a fractional width
0.10 for the interval spacing.  Figure~\ref{fig:lpfeat} shows a
greyscale coding of the number of trials registering at each
location, the position of peaks or spans of flats. The greyscale
levels, marked at the top of each graph, differ between plots;
they show the relative frequency of the location starting at 5\%
of the maximum count.  The counts are higher for flats because
they overlap while the point features do not register precisely
the same location.  The mode and/or anti-mode locations are drawn
as in Figure~\ref{fig:bimodal}.  The counts are for detected
features and a contour line is added enclosing 90\% of those that
pass any of the tests, model or run or excursion, at the
recommended level.

The top row of Figure~\ref{fig:lpfeat} shows that the peaks track the
anti-mode.  The detected position becomes more variable, with a wider
greyscale envelope, as the separation decreases or width increases,
but extends until the mode disappears.  The detected peaks are
insensitive to the change in draw size, except that one sometimes
appears in the first mode if it is much larger than the second.
Testing does a good job of rejecting peaks in these transitional
regions, as the contour area stays close to the anti-mode and does
not widen as much.  Peaks become significant near the default setup,
for separations greater than 2.8 or standard deviations below 1.2.
This is equivalent to a rise of 0.003.

Detected flats in the middle row of Figure~\ref{fig:lpfeat} appear
in the modes, but not always cleanly.  For example, one flat covers
both for separations below 2.6, and in half the trials a flat spans
both when the second draw's standard deviation is between 1.2 and
1.8.  Here the rise is too small, below 0.002, to interrupt the
ripple specification.  The plot can be misleading, as the sum of
the spans can conceal multiple short flats.  Small flats often
appear within the anti-mode, visible in all three graphs as fingers
that disappear when the peaks become significant, for example when
the separation is between 3.0 and 3.7, or for all draw sizes.  Under
these conditions the peaks are apparently wide enough to support a
minimum length flat.  This is why the detector collapses nearly
equal points to one.  Passing flats identify both modes when the
separation is above 3.0, the draw sizes are within 150 of each
other, and the standard deviation 1.0 or less.  The rise at these
points is 0.0052--0.0077.  Outside these ranges they identify one,
the merged draws at small offsets or the larger draw when
unbalanced, or the first mode as it begins to absorb the less well
defined second variate as its standard deviation grows.

We discussed in Section~\ref{sec:di} that changepoints will mark
neither modes nor anti-modes directly, responding to the transition
between them.  This can be seen in the bottom row of
Figure~\ref{fig:lpfeat}.  The greyscale bands lie to the side of
the anti-mode line and they avoid the modes.  This is clearest in
the draw size variants.  The sensitivity of the detectors compared
to the features is mixed: in the offset variants changepoints start
appearing at a mean of 3.2 and grow stronger with larger
separation, and for standard deviations below 0.9, rises of 0.007.
The changepoints respond most strongly to the initial and final
tails where the darkest greyscale levels lie along the left and
right sides of the graphs and the vertical edges are wide.  Indeed,
the broad greyscale bands show the changepoint positions are much
less stable than the peaks.

\begin{figure}[ht]
\centering
\begin{minipage}[t]{\textwidth}
\includegraphics[width=\textwidth]{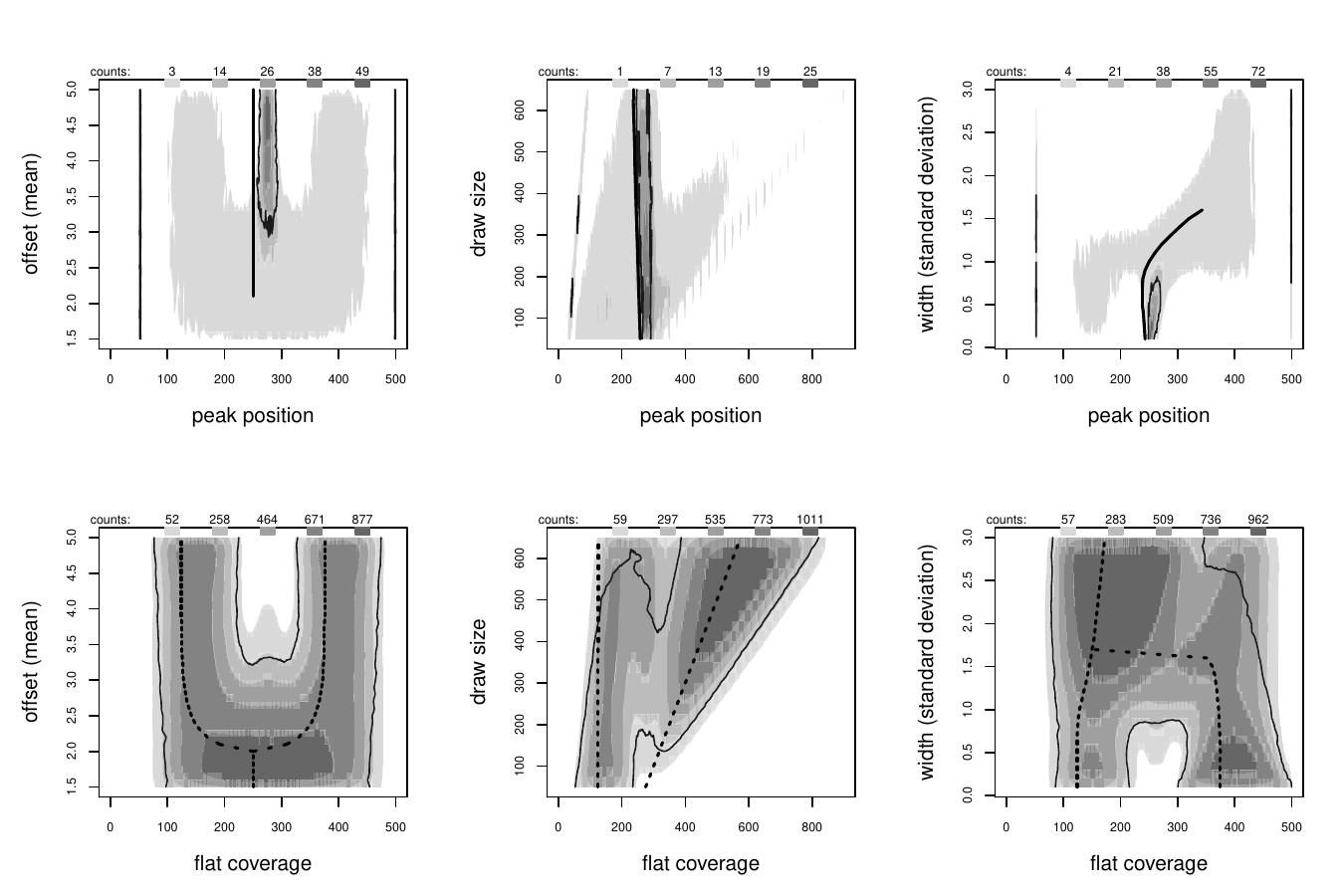}
\caption{\label{fig:diwfeat} Peak (top) and flat (bottom) positions
in the interval spacing, with same markings used in
Figure~\ref{fig:lpfeat}.
}
\end{minipage}
\end{figure}

We see a larger dispersion in the peak position in the interval
spacing (Figure~\ref{fig:diwfeat}) at separations below 3.5
or widths above 1.0, and spurious peaks appear inside the modes at
all separations.  There are also peaks that appear in the initial
and final tails for all variants.  Tests successfully screen these
extra features and pass peaks for the same variants as the
low-pass spacing, with a separation of 2.7 or standard deviation
below 1.2.  This corresponds to a rise of 0.0025. All draw size
variants meet this rise because of the larger offset, and accepted
peaks follow the anti-mode.  The passing contour is wider,
however, and these peaks are less stable.  Their positions appear
shifted to the side of the anti-mode because of the indexing
convention for $ D_{i,w} $, by half the interval width.  This
shift is also visible with the flats.  The interval spacing has
somewhat more flats than the low-pass spacing, despite its
roughness.  They span the same range of indices.  A larger range
of variants accepts the features because the rough signal make
it harder to replicate them in the excursion test.  The contour
line follows the outer greyscale level, implying that most that
are found will pass testing, while the detected peaks extend
far beyond the contour, meaning testing rejects many of the
marginal cases.

All three features demonstrate that the data's spacing reflects
its modality.  Peaks mark anti-modes, flats modes, and
changepoints the transitions between them.  The tests reject
marginal or transitional features, keeping to the ideal position
when the two variates are well-defined with rises greater than
0.003 or to the dominant variate when they are unequal.  When
used as an overall modality test, we expect to see a shift in
the probability around this value.

\begin{figure}
\centering
\begin{minipage}[t]{\textwidth}
\includegraphics[width=\textwidth]{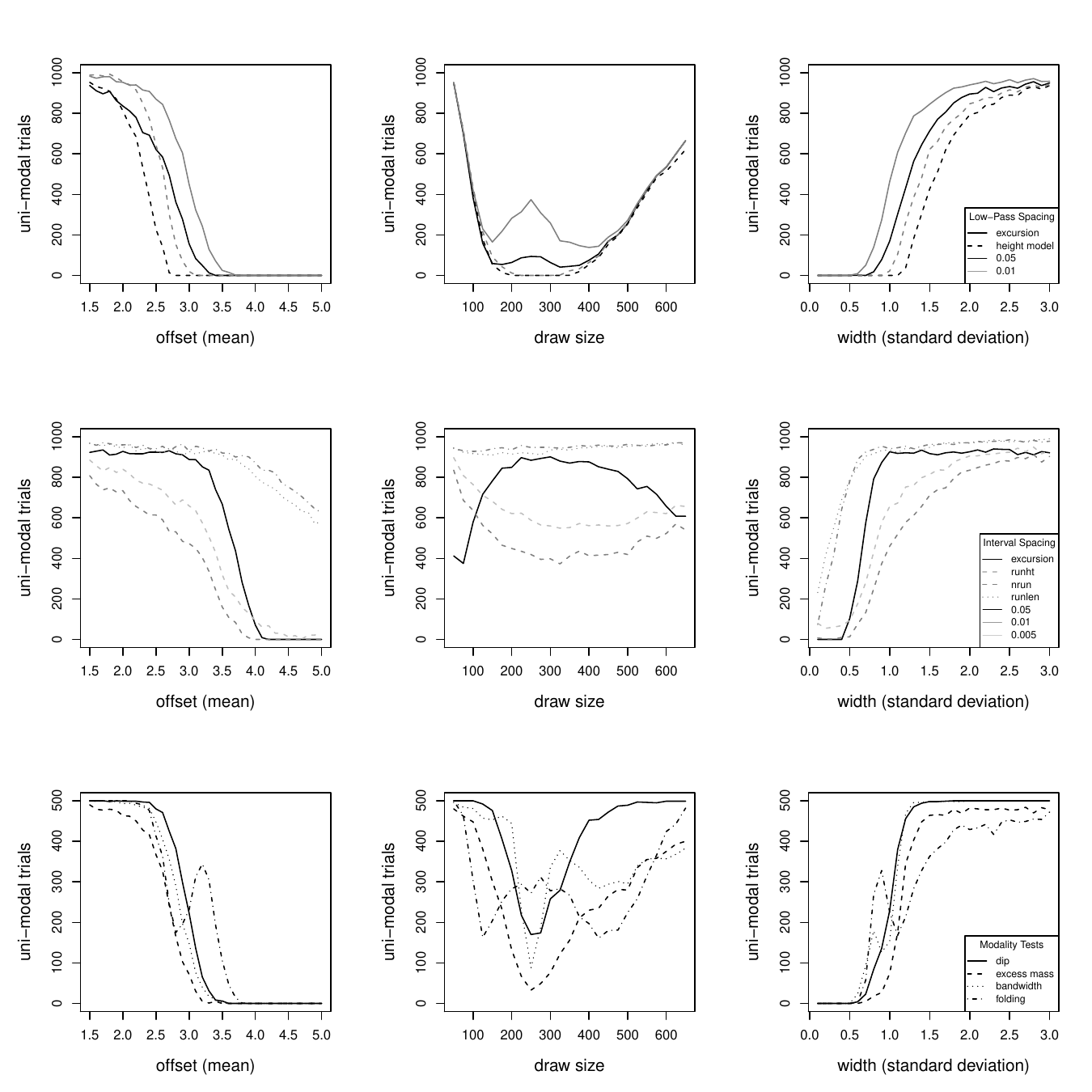}
\caption{\label{fig:pmodal} Rejection rates of uni-modality for peaks
in the low-pass spacing (top) and interval spacing (middle), and using existing
tests (bottom).
}
\end{minipage}
\end{figure}

And we do.  Figure~\ref{fig:pmodal} counts the number of trials
of the one thousand repetitions where a test accepts a peak,
rejecting uni-modality.  The top row presents the low-pass
spacing tests and the middle those in the interval spacing.
The height model accepts peaks at a separation 0.4 smaller than
the excursion test at the same passing level, a difference in
the rise of 0.0021 -- 0.0033.  As discussed in
Section~\ref{sec:models}, the confidence interval about the
0.95 quantile is large enough to recommend using the 0.01 level
for the height model, and this brings both curves together,
choosing multi-modality at a separation of 2.6 or a rise of
0.0023.  The rejection rate of the excursion test ramps down
for smaller offsets rather than transitioning sharply,
allowing marginal peaks to pass.  At the 0.01 acceptance level
the ramp disappears but the offset must be larger, so the
sensitivity is lower.  We see the same general trends with the
standard deviation variants: the height model accepts peaks at
widths 0.35 larger, a difference in the rise of 0.003, with
the excursion test less sensitive by 0.2 (0.001) at the
recommended levels.  Both the separation and width variants
consistently resolve the bi-modal setup.  Complete separation,
with a minimal uni-modal or false negative rate, occurs at a
separation of 3.1 -- 3.4 (0.0063 -- 0.0103), for standard
deviations below 0.8 -- 1.0 (0.0078 -- 0.0052), and draws
between 150 and 425 (0.0117 -- 0.0073).  The rapid rise for
the draw variants to the left and gradual change to the right
are similar when considered as the ratio of the imbalance.
If we had used an offset of 3.5 for these variants only draws
below would 100 would not be fully resolved, implying a
sensitivity to the total sample size.  At 3.1 the excursion
test would not fully distinguish the setup.

The interval spacing tests in the middle row are not as good at
judging the modality.  The excursion requires larger rises, by 0.01,
which correspond to separations increased by 0.9 to 3.6 or widths
decreased by 0.6 to 0.7.  The transition is sharper.  It finds
balanced draws harder to distinguish despite their larger rise.
The run height test is much more sensitive, but it is not decisive.
It accepts peaks even when there is no anti-mode, with the rejection
rate steadily increasing until it switches near the default setup.
Evaluating the test at the 0.005 level helps lower this bias but
does not eliminate it; the test has a high false positive rate.  It
is insensitive to draw sizes above 150, again implying that the
total sample size has an effect.  Changing the default offset by
$ \pm 0.2 $ shifts the rejection rate by $ \mp 20\% $ for this and
the excursion test.  The runs count behaves the same as the longest
run test.  They resolve the standard deviation variants, needing a
width 1.0 smaller than the excursion test, (rise $ +0.016$).
They cannot resolve the offset variants for separations smaller
than 4.0 and respond very slowly to larger values.  This is above
the default offset and they cannot resolve the draw size variants.
If the acceptance level were loosened to 0.05 their responses
would shift vertically downward by 20\%.  This is similar to the
change in the run height curve, but larger.

The excursion test is the only reliable check of peaks in the
interval spacing.  The runs height test would be were it not
biased towards accepting marginal peaks, even at the 0.005 level.
A positive result from the longest run or runs count will
provide strong evidence of multi-modality but should mostly
confirm the excursion test.

Compare these results against existing multi-modality tests.
The Hartigans' dip test \citep{hartigan85} uses a test statistic
based on the difference of the data's distribution to a uniform
variate.  The Silverman critical bandwidth test
\citep{silverman81} calculates the bandwidth or kernel size of
the Gaussian low-pass filter needed to generate a given number
of modes for draws from a uniform variate and compares this
critical value against the number actually present.  We do not
use any of the corrections to the critical value that have been
proposed to compensate for the many peaks the uniform null
distribution creates.  The excess mass statistic, with
corrections per \citep{ameijeiras16}, compares the density above
and below a threshold varying between 0 and 1, using the value
producing the largest difference as the test statistic.  The
critical bandwidth and excess mass tests take as a parameter
the number of modes in the data, and so can identify how many
there are.  The folding test \citep{siffer18} is designed to
detect multi-modality in data with more than one dimension.
It compares the variance of the difference between the data
and some fixed point against the raw variance, with the fixed
point estimated by the data's second and third moments in
the one dimensional case.  None of these four tests provide
a location for modes.

The bottom row of Figure~\ref{fig:pmodal} shows the uni-modal
pass rates over half of the trials of each bi-modal variant,
judging at the 0.05 level.  The excess mass test is the most
lenient, reliably rejecting uni-modality when the separation
is 2.7 (rise 0.0026). The critical bandwidth recognizes
bi-modality at a separation of 2.8 (0.0034), and the dip test
at 3.0 (0.0052).  For the width variations the excess mass
test reliably rejects uni-modality at a standard deviation of
1.1 (0.0036) and the Silverman and dip tests at 0.8 (0.0078).
These tests are a little less sensitive than the spacing.
All transitions are sharp.  The folding test has a kink in
its results because it reports that the data is uni- or
multi-modal and assigns a probability to the judgement.  The
graph plots the uni-modal pass rate until the switch occurs,
and then the multi-modal failure rate after.  The switch
happens to fall at the default setup.  The folding test is
closest to the excess mass result until the switch and then
becomes the most conservative of all, requiring a separation
of 3.5 (0.0119).  These tests resolve only balanced draws.
The critical bandwidth and excess mass tests partially
resolve unbalanced draws, behaving the same as the low-pass
tests were the default offset to be 3.1.  The folding test
does not react to the imbalance.  An offset of 3.5 would
improve the rejection rate, even in the folding test, but
the interaction of the three parameters near the critical
rise is complicated and produces marginal results.

In summary, the low-pass tests are slightly more sensitive than
the excess mass, the most sensitive of the existing tests, and
handle unbalanced draws better.  The interval excursion test
needs a larger peak, but behaves better than the runs tests.
These are notably slow to respond to changes in the rise, a
problem which afflicts none of the existing tests.  The run
height permutation test also suffers from a slow transition
and is a little less sensitive than the dip test.  If its
transition were to sharpen it would be equivalent to the
low-pass tests.

As a second check on the ability of a test to detect
multi-modality, we can present it with data drawn from a single
variate and count in how many trials the test claims the draw
is not uni-modal; the rate should match the acceptance level.
Of 1000 trials of draws of 300 points from a uniform or normal
distribution, Table~\ref{tbl:unimodal} presents the fraction
that have a peak passing at least one test.  Rates that match
or better the 0.05 or 0.01 level are underlined, while rates
at the next higher level have wavy underlines.  We see that
the dip and critical bandwidth tests perform as expected with
uniform draws, because they are calibrated for them.  They do
not misclassify any of the normal draws.  The folding test
does as well.  The excess mass test is based on normal
variates and matches their rate, but passes too many uniform
draws.

\begin{table}
\centering
\caption{\label{tbl:unimodal} Uni-Modal Rejection Rates}
{\small
\begin{tabular}{rlrccrcc}
 & && \multicolumn{2}{c}{normal N(0,1) draws} &&
      \multicolumn{2}{c}{uniform U(0,1) draws} \\
 & && 0.05 level  & 0.01 level   && 0.05 level   & 0.01 level \\
\multicolumn{8}{l}{existing tests} \\
 & Hartigans' Dip &&
     \margv{0.000} & \margi{0.000} && \margv{0.049} & \margi{0.013} \\
 & critical bandwidth &&
     \margv{0.000} & \margi{0.000} && \margv{0.060} & \margi{0.005} \\
 & excess mass    &&
     \margv{0.023} & \margi{0.001} && \margv{0.231} & \margi{0.088} \\
 & folding        &&
     \margv{0.000} & \margi{0.000} && \margv{0.050} & \margi{0.011} \\
\multicolumn{8}{l}{low-pass spacing} \\
 & height model   &&
     \margv{0.007} & \margi{0.002} && \margv{0.983} & \margi{0.970} \\
 & peak excursion &&
     \margv{0.016} & \margi{0.007} && \margv{0.936} & \margi{0.878} \\
\multicolumn{8}{l}{interval spacing} \\
 & peak excursion &&
     \margv{0.079} & \margi{0.066} && \margv{0.225} & \margi{0.046} \\
 & run height     &&
     \margv{0.271} & \margi{0.093} && \margv{0.764} & \margi{0.337} \\
 & run count      &&
     \margv{0.088} & \margi{0.031} && \margv{0.312} & \margi{0.092} \\
 & longest run    &&
     \margv{0.100} & \margi{0.024} && \margv{0.284} & \margi{0.071} \\
\end{tabular}
}
\end{table}

To check the spacing tests, we run the same draws through a Kaiser
filter of width 60 (20\%) or use an interval of 45 (15\%).  The
low-pass tests correctly screen peaks from the normal draw, with
the excursion test rates resembling the excess mass test.  A
quarter of the trials contain detected peaks, so the screening by
the tests is successful.  The interval spacing has a detected peak
in 90\% of the trials and screening is less successful.  Only the
runs count and longest run tests come close to the desired 0.01
rate, but they still have some false positives, moreso at the 0.05
level.  No tests succeed in the uniform trials, which generate
many more peaks, 1.9 per trial in the low-pass and 4.1 in the
interval spacing; this is why the height model does not use it as
the null distribution.  Checking the run height permutation test
at the 0.005 level has a large effect, lowering the rejection
rate to 0.055 for the normal draws and 0.220 for the uniform, but
this reflects its bias towards accepting marginal peaks.

If we were to use less smoothing, using a filter size of 45 (15\%)
and interval of 30 (10\%), only the height model would reject
uni-modality for the normal draws at the correct rate.  The
low-pass excursion and longest runs tests would have a rejection
rate of 0.028 at the 0.01 level.  The data-driven tests still seem
sensitive to the filtering that has been done.

\section{Test Samples from the Literature}
\label{sec:litds}

Modality testing has generated a rich literature and many papers
have defined ideal samples to illustrate or evaluate the many
approaches explored.  They tend to be difficult cases, with small
modes against a large background or with little separation
between modes.  Because the examples are built from known
variates, mostly combinations of Gaussians, we can solve for the
expected spacing numerically and compare the spacing results to
determine which modes and anti-modes are found, the accuracy of
the feature's placement, and the stability over repeated draws.
We generate each sample 400 times, run the feature detectors
and tests on the low-pass and interval spacings, and gather the
results.  We use the default detector parameters and accept and
count features at the recommended levels.

The samples are usually named as in their source.  \citep{ameijeiras16}
defines the A series, replacing `M' in the name by `A'.  Draws A1
through A10 are unimodal, although often built from two or three
variates placed to create shoulders and extend tails.  A11 through
A20 are bi-modal, sometimes also created with a third variate.  A21
through A25 are tri-modal.  The C examples, from
\citep[eq. (3.1), (3.2), (3.3)]{cheng99}, are uni-modal with a second
normal added as a shoulder.  The D series is found in
\citep{mukhopadhyay16}; we do not include the uni-modal examples.  The
F samples come from \citep[Table 2]{fisher01}, with F1 labeled
Contaminated, F2 Small Blip, F3 Asymmetric Bimodal, F4 Trimodal, and
F5 Four Modes. G1 \citep{duembgen08} is a tri-modal setup using draws
from two normals and a gamma distribution.
\citep[Section 3]{holzmann08} defines the H samples as alternatives
a, b, c, and d.  H1, H2, and H3 use two normal variates, H4 two
$ t $ draws.  M1, M2, and M3 are bi-modal normal draws defined in
\citep[Table 1]{minnotte97}, and M4 is a tri-modal example from
Table 2.  The N series is built from multi-modal draws
\citep[Section 3.1]{davies04} with the number being the mode count.
Section 3.7 of the paper defines the discrete P1 sample using
Poisson variates.  The numbering of the W series follows
\citep[Table 1]{marron92}.  W4 and W5 are uni-modal, W6, W7, and W8
bi-modal, W9 tri-modal, and W10 and W12 penta-modal.  The X series
are the multi-modal entries in \citep[Table 1]{xu14}, numbered in
order starting with the fourth row. X1 through X4 are bi-modal, the
others tri-modal.  There are not many complicated examples in this
set, and Appendix~\ref{app:ksample} defines additional quad- and
penta-modal samples, not all using normal draws.  Two are discrete.

\begin{figure}
\centering
\begin{minipage}[t]{\textwidth}
\includegraphics[width=\textwidth]{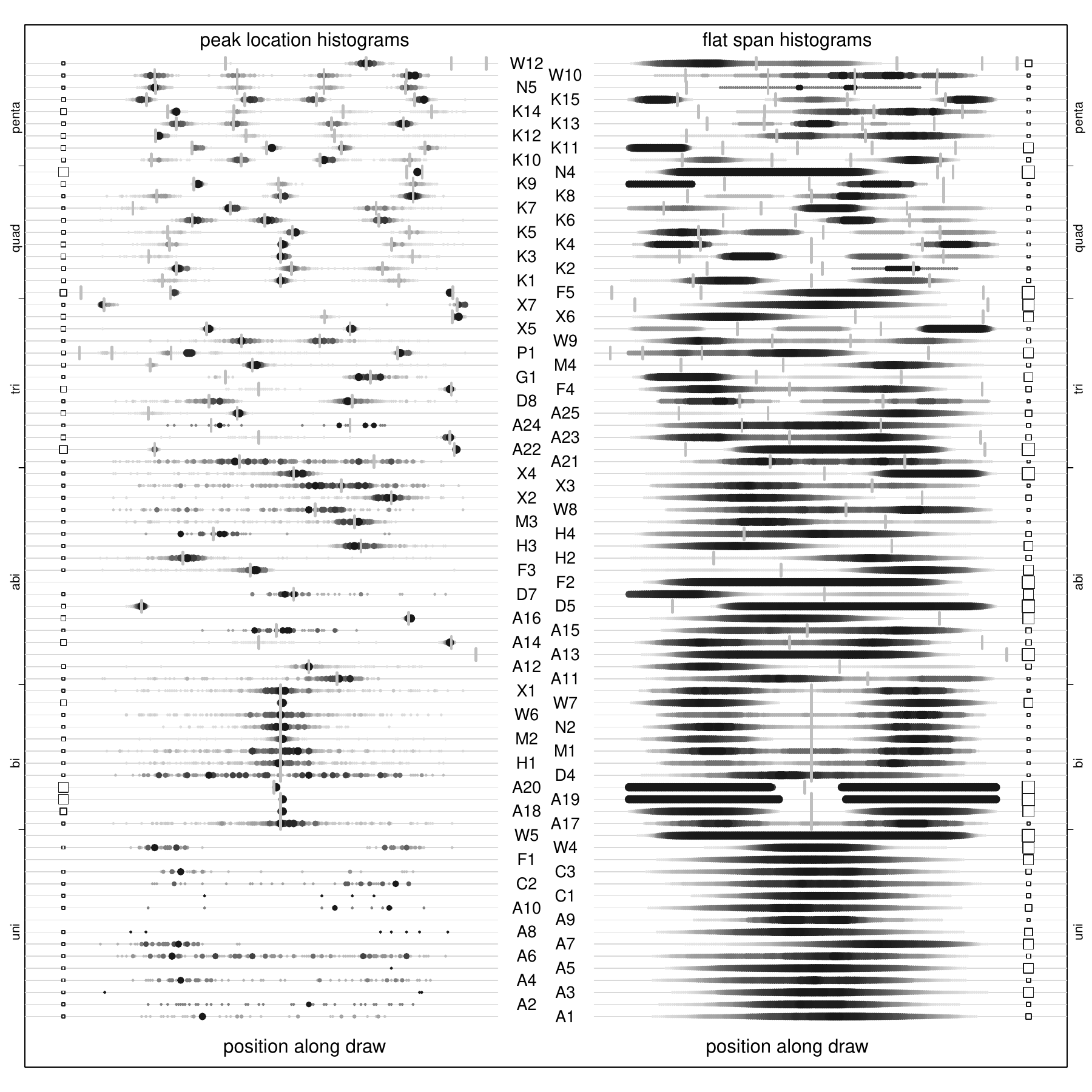}
\caption{\label{fig:dslp} Low-pass peak locations (left half) and flat
 spans (right) over trials of test samples from the literature.  Grey
 level and point size represent fraction of counts, size of box at edge
 indicates maximum count.
}
\end{minipage}
\end{figure}

A greyscale histogram for each sample, equivalent to a slice through
one of the graphs in Figure~\ref{fig:lpfeat}, shows the accuracy and
stability of the passing features.  The greyscale levels and size of
the dots in Figure~\ref{fig:dslp} are consistent between samples and
represent the fraction of the maximum count in that bin. The maximum
counts are shown by the size of the the boxes at the ends, which
increases for each multiple of 50.  The $ x $ axis has been scaled
by the size of each draw.  The figure groups samples by the number
of modes they contain, with `abi' standing for
``asymmetric bi-modal''.  Ticks mark anti-modes, and peaks are
accurately placed if they lie there.  Features are stable if they
have black dots or bars with abrupt edges that do not trail away.
They are repeatable if the box to the side is large.

The left half of Figure~\ref{fig:dslp} counts accepted peaks, the
right flats.  For peaks the largest maximum count is 365 (quad-modal
sample N4) and for flats 443 (quad-modal F5).  That the flat count
is above the number of trials means that there are overlapping
flats.  The bars cannot distinguish between a single long flat or
several adjacent, especially when gathered over all the trials.

Bi-modal samples A18, A19, A20, and W7 have the best peaks, showing
tight dots at the anti-modes and the largest boxes.  They are
accurate, stable, and repeatable.  They also have large rises,
0.020, 0.008, 0.018, and 0.076 respectively.  These are above the
detectable limits found in the bi-modal analysis.  Asymmetric
bi-modal samples A14, A16, and D5 are tight and placed correctly,
but appear less often.  They have similar rises.  Most of the other
abi samples are poorly defined or inaccurately placed, with small
maximum counts reflecting the spread-out positions.  Many of these
samples involve a small draw to the side of a much larger, and its
larger variation creates the instability.  Tri-modal sample X5 is
resolved, while the peaks mark one anti-mode in A22, A23, A25, and
F4.  These peaks are repeatable, but not as much as the bi-modal
samples.  Three samples, A13, F5, and W12, have anti-modes that
match no significant peaks.  They are lost in the tails, despite
having rises well above those that could be resolved in the
bi-modal variates.  Quad- and penta-modal samples K6, K13, and
K15 are fully resolved, the others partially.

We can quantify the accuracy of a peak's placement using either
the distance between it and the anti-mode, taking the inter-quartile
range for robustness, or by counting trials where the distance is
small, within 1\% of the draw size.  Either measure is only roughly
inversely related to the rise.  The largest rises do have the best
accuracy and the smallest the worst, but there is no consistent
trend in-between.

The samples with the most repeatable peaks are also stable and
accurate. These include the bi-modal A18, A19, A20, and W7, the
right anti-mode of tri-modal A22, and two in F5.  The peaks in N4
fall between two close anti-modes.  Many other peaks such as A14
and D5 are less stable but still accurate.

Flats, on the other hand, do not have stable endpoints.  A19, A20,
and the left mode of K9 are the exceptions; most bars smoothly
trail away, such as A13.  There does not seem to be a relationship
between the rise and the endpoint stability.  Nor does it affect
the overlap of anti-modes, which happens in most of the abi and
tri-modal sets and at a background rate in several bi-modal
samples.  Bi-modal D4 cannot be separated, but its rise is 0.0005,
which we saw in the bi-modal variants is not enough to distinguish
the modes.

\begin{figure}
\centering
\begin{minipage}[t]{\textwidth}
\includegraphics[width=\textwidth]{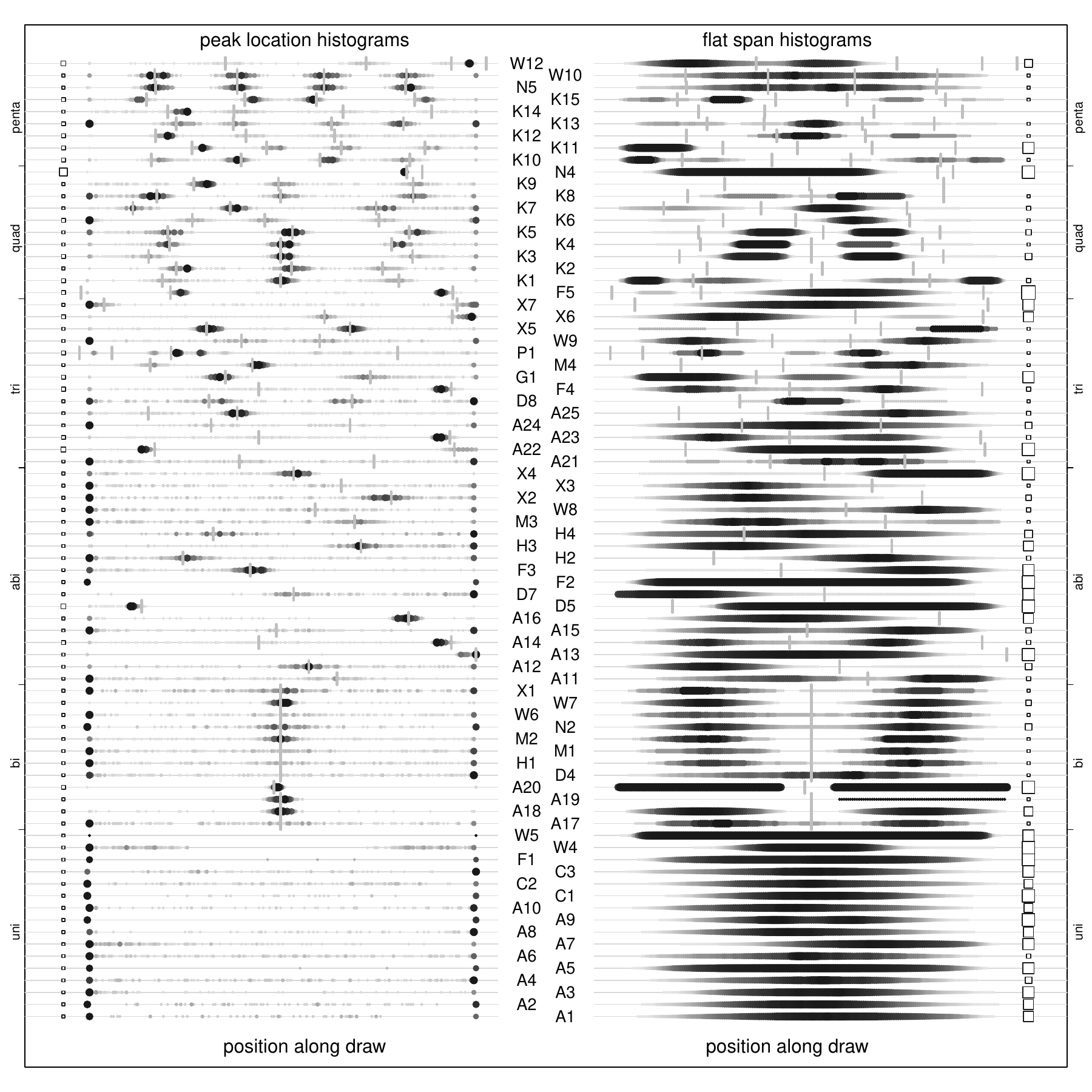}
\caption{\label{fig:dsdiw} Accepted feature counts in the interval spacing.
  Same greyscale and point size encoding as in Figure~\ref{fig:dslp}.
}
\end{minipage}
\end{figure}

Figure~\ref{fig:dsdiw} plots the interval spacing features from
the same draws.  Peak positions are in general less stable,
especially in the abi samples.  There is a high background rate,
where peaks scatter over all indices, that speaks to the roughness
of the interval spacing.  This is best seen in the uni-modal
samples.  Noise in the tails creates single point outliers at
the edges of the graph.  Tests do accept the peaks at the edge
of A13 and the combination of the last two in W12, so the
interval spacing can resolve features in the tails that the
low-pass spacing cannot.  Unstable peaks in the low-pass spacing
disappear, such as H1, M1, X3, and A21.  The interval spacing does
a better job with the penta-modal samples.  Peak positions are
less repeatable, with a maximum count of 290 (sample N4).  Only
three samples have maximum counts above 100, compared to twenty
in the low-pass histograms.  Flats in the bi-modal samples behave
similarly in both spacings, but there are more interval flats in
the complicated setups, particularly the quad-modal. There are
some strange results, for example the easily separated A19 has
flats only in the right mode.  The maximum flat span count is
473 (sample F5), higher than the low-pass spacing.

M1, H1, and N2 have the same setup as the default bi-modal variant,
with balanced draws, unit standard deviation, and a separation of
3.0.  N2 draws 250 points for each variate, M1 and H1 100.  X1 has
a larger separation, 4.0, but also a larger standard deviation, 1.25,
and the two changes mostly cancel.  The combs W10 and N5 have the
same setup, differing in their draw size, 30 for each variate in
W10 and 50 for N5. The H1 and M1 peaks and flats have similar
histograms in both spacings, so 400 trials seems to be enough for
reliable results.  The larger N2 sample produces peaks that are
more stable, which is also true for the W10/N5 pair.  X1 peaks fall
in-between H1/M1 and N2.  The accuracy is the same.

Peaks in both spacings match.  There are fewer in the low-pass
spacing but they have more stable positions.  The number of peaks
in the interval spacing that lie close to the anti-mode, within
1\% of the sample, is half that in the low-pass spacing,
0.21~matches/trial vs. 0.44, which means they are less accurate at
locating the anti-mode.  Peaks in the interval spacing have more
variability in their position due to the small suppression of high
frequency components.  This roughness also creates more peaks,
with twice as many detected, 3.4 per trial averaged over all
samples, than in the low-pass spacing, 1.6.  Testing rejects more
of the interval spacing peaks and the number of significant peaks
is the same in both spacings, 1.0 per trial.  The consistency of
the peak positions changes during testing.  68\% of the detected
low-pass peaks align within 2\% of the sample to an interval
peak, but in the other direction 34\% do.  Testing equalizes the
rate so that a third of the passing peaks in either spacing have
a match.

Low-pass peaks match anti-modes better.  Let 
$ (N_{am} - N_{m}) / (N_{am} + N_{m}) $
define the selectivity, where $ N_{am} $ is the number of peaks
over all trials that are closer to an anti-mode than mode and
$ N_{m} $ the count associated with the mode.  The metric will
be $ +1 $ when all features match the anti-modes, even if
loosely, and $ -1 $ when they align only with modes.  Over all
examples the median selectivity is $ +0.84 $ for the accepted
low-pass peaks and $ +0.35 $ for the interval.  The high
background rate in Figure~\ref{fig:dsdiw} can align to modes
and lowers the selectivity. 11 anti-modes have no nearby
low-pass peaks in any trial, and 7 have no interval peaks,
while 121--125 are found at least once.

Flats match better.  There is no difference in the number found
per trial, 1.25 in either spacing, and testing works equally but
strictly, passing one fifth, 0.25.  Defining matching flats as
those that overlap for 70\% of their length, half of the detected
features and a third of the passing align between spacings.
Instability in the endpoints from the roughness of the interval
spacing combines with the low acceptance rate to reduce the pool
of potential matches and therefore the rate.  Testing selects
flats covering the modes. 0.69 flats/trial include a mode, while
0.14 cover an anti-mode, 0.40 include neither, and 0.03 both.
The passing flats only span a mode, or occasionally lie to the
side and include neither mode nor anti-mode in 10\% of the trials.

Changepoints mark neither mode nor anti-mode, nor are they stable
over trials. They somewhat favor modes but not strongly, with a
selectivity of $ -0.13 $; changepoints identify transitions in
the spacing.  We get in general many changepoints, including in
the uni-modal samples which generate 1.5 per trial on average,
which is the same rate as in bi-modal samples. Many changepoints
trigger at the leading and trailing tails of the spacing, in the
arms of the `U'.  40\% of the changepoints lie within 5\% of the
draw from either end, and 65\% within 10\%.  Several of the
algorithms apparently have a burn-in interval of 15 points for
data statistics before they being to look for changes, as 83\% of
the samples (68 of 81) generate their first changepoint at this
index.  Histograms of their positions are generally not
concentrated, even for strongly defined samples like A19 and A20.
The average extension of the histogram around a local peak is
19.2 data points, with a strong tail and 90\% confidence interval
extending out to 58.3 points.

\section{Conclusion}
\label{sec:conc}

The overall evaluation of the spacing tests can get lost in the
numbers and discussion.

The low-pass filter models are accurate, but degrade at the edges
of the model parameter space.  The accuracy of the peak height
model becomes suspect for $ n \le 70 \text{ or } n \ge 500 $, and
for $ f_{lp} \le 0.075 \text{ or } f_{lp} > 0.30 $.  For flat
length models the bounds are $ n \le 60 $ and $ f_{lp} \le 0.075 $.
Peak heights should be tested at the 0.01 level, while flats can
be accepted at 0.05.

Peaks are reliably found and judged significant when the expected
spacing rises by 0.003, equivalent to a separation of two normals
by 2.8 or a standard deviation in one of 1.2 or for all draw
sizes.  A larger rise of 0.007 is enough to fully distinguish the
variates.  Smaller rises are detected but the position becomes
less stable as the rise decreases.  A rise of 0.002 is enough to
break the ripple specification and generate separate, passing
flats in each mode. This is equivalent to a separation of 2.6 or
standard deviation of 1.2.  The resolution is the same for both
spacings.

There is only a rough relationship between the increase in
spacing in the literature samples and the features detected.
The accuracy of the peak placement to the anti-mode improves
with the rise, but the stability does not.  Peaks in the
low-pass spacing align better to the anti-modes than those in
the interval spacing.  More peaks are found in the interval
spacing, but testing equalizes the count of significant
features.  This comes at a cost in agreement of the position
of the two kinds of peaks, which is lower between accepted
than detected features.  Flats in the two spacings match,
with equivalent detection and acceptance rates, coverage of
modes, and position.

Changepoints do not mark features directly.  They fall to the
side of the anti-modes, not on them.  The various detectors are
noisy, which the majority voting algorithm does not completely
eliminate.  This is seen in the number of changepoints
identified and the stability of their position.  The detectors
have a high background rate, finding as many points in the
uni-modal literature samples as the bi-modal, mostly because
the detectors are sensitive to the strong changes in the
initial and final tails.

Overall, the presence of peaks, not flats, signals multi-modality
in the data.  Flats place requirements on the number of points
coming from a mode, are most obvious when there are dramatic
changes in the spacing that sharply sets the edges, and are
sensitive during analysis to erosion of the ends by the filters
or noise in the data.  A secondary effect, the average spacing
within the flat, has not gotten much discussion here.  It
depends on the distribution's parameters, such as the standard
deviation of a normal or rate of an exponential, but without a
model of the data we cannot turn the values into concrete
statements about the variates.  There are hints in these
results that the spacing can respond to buried changes, for
example in samples C3 and A7, where the overall distribution
is smooth but the transition between variates creates a small
rise in the spacing that sometimes appears in individual trials.
The low-pass peak tests are sensitive to modality changes, with
sharp transitions as the parameters change.  They do a little
better than the excess mass test.  The interval spacing tests
are less decisive and need more distinct modes.  None of the
spacing tests do well with a null distribution from a uniform
variate.  All three runs tests need tight acceptance levels to
minimize false positives.

Some open questions remain.  Do the results extend to large
data sets, with more than 10 thousand points?  The spacing in
this case is very small and generally smooth, so little
filtering is needed.  But these conditions --- small filters
and large data sizes --- are just those where the model
accuracy might not be good.  Perhaps the excursion and runs
tests will do better under these conditions.  Similarly, do
the results hold when we change detector parameters?  The
roughness of the interval spacing might require a relaxed
ripple specification for flats, but the test performance under
these conditions is not known.

Filtering, explicitly or implicitly through the interval
spacing, smooths over the steps created by discrete or
strongly quantized data and the spacing analysis can study
the modality of these samples. This is important, because
limited precision in the data, for example if taken to two
or three decimal places, will appear in the spacing. Low-pass
filtering does better under these conditions, where the
interval spacing is generally too rough and changepoints tend
to trigger on each step.  Features get lost in the initial
and final tails of the data, where the spacing increases
quickly. The interval spacing will have more success in
these cases, as it responds more quickly to local changes.

\section{Software and Supplementary Material}
\label{sec:sw}

The R package \rpkg{Dimodal} \citep{kreider25e} provides the reference
implementation of the feature detectors and tests \citep{kreider25c}.
The data files used for the results are provided as supplementary
material, as are the scripts that generated them and created the
figures.

\section*{Disclosure of Interest and Funding Statements}
There are no competing interests to declare.  No funding was received
for this work.

\bibliographystyle{tfcad}
\bibliography{dmodal}

@preamble{ " \newcommand{\noop}[1]{} " }

@ARTICLE{ameijeiras16,
  author = "Jose Ameijeiras-Alonso and Rosa M. Crujeiras and Alberto Rodr\'{i}guez-Casal",
	title = "Mode Testing, Critical Bandwidth and Excess Mass",
	journal = "Test",
	volume = 28,
	number = 3,
	year = 2016,
	month = sep,
	pages = "900-919"
}

@ARTICLE{cheng99,
  author = "Ming-Yen Cheng and Peter Hall",
	title = "Mode Testing in Difficult Cases",
	journal = "The Annals of Statistics",
	volume = 27,
	year = 1999,
	pages = "1294--1315"
}

@ARTICLE{davies04,
  author = "P. Laurie Davies and Arne Kovac",
	title = "Densities, spectral densities and modality",
	journal = "The Annals of Statistics",
	volume = 32,
	number = 3,
	year = 2004,
	pages = "1093--1136"
}

@ARTICLE{duembgen08,
  author = "Lutz D{\"u}mbgen and G{\"u}nther Walther",
  title = "Multiscale Inference about a Density",
  journal = "The Annals of Statistics",
  volume = 36,
  number = 4,
  year = 2008,
  pages = "1758--1785",
  note = "doi: 10.1214/07-AOS521"
}

@ARTICLE{fisher01,
  author = "N. I. Fisher and J. S. Marron",
	title = "Mode testing via the excess mass estimate",
	journal = "Biometrika",
	volume = 88,
	number = 2,
	year = 2001,
	pages = "499--517"
}

@ARTICLE{hartigan85,
  author = "J. A. Hartigan and P. M. Hartigan",
	title = "The Dip Test of Unimodality",
	journal = "The Annals of Statistics",
	volume = 13,
	number = 1,
	year = 1985,
	pages = "70--84"
}

@ARTICLE{harris78,
  author = "Fredric J. Harris",
  title = "On the Use of Windows for Harmonic Analysis with {D}iscrete
           {F}ourier {T}ransforms",
  journal = "Proceedings of the IEEE",
  volume = 66,
  number = 1,
  year = 1978,
  month = jan,
  pages = "51--83"
}

@ARTICLE{holzmann08,
  author = "Hajo Holzmann and Sebastian Vollmer",
	title = "A likelihood ratio test for bimodality in two-component mixtures ---
           with application to regional income distribution in the {EU}",
	journal = "AStA Advances in Statistical Analysis",
  volume = 92,
	number = 1,
	year = 2008,
	pages = "57--69"
}

@ARTICLE{kreider23a,
  author = "Greg Kreider",
  title = "Expected Spacing",
  journal = "Communications in Statistics - Theory and Methods",
	volume = 53,
	number = 23,
  year = 2023,
  pages = "8286--8296",
  note = "doi: 10.1080/03610926.2023.2281265"
}

@UNPUBLISHED{kreider25a,
  author = "Greg Kreider",
	title = "Using Spacing to Detect Multi-Modality",
	year = "2025",
  note = "https://arxiv.org/abs/2608.18228"
}

@UNPUBLISHED{kreider25c,
  author = "Greg Kreider",
	title = "Modality Analysis via Spacing with the {D}imodal Software Libraries",
	year = "2025",
  note = "https://arxiv.org/abs/2607.06722"
}

@UNPUBLISHED{kreider25d,
  author = "Greg Kreider",
  title = "Runs and Bootstrap Tests for Signal Feature Significance",
  year = "2025",
  note = "https://arxiv.org/abs/2607.09913"
}

@MANUAL{kreider25e,
  title = {{\bf Dimodal}: Spacing Tests for Multi-Modality},
	author = "Greg Kreider",
	year = "2025",
	url = "https://www.primachvis.com/data/Dimodal\_latest.tar.gz"
}

@ARTICLE{marron92,
  author = "J. S. Marron and M. P. Wand",
	title = "Exact mean integrated squared error",
	journal = "The Annals of Statistics",
	volume = 20,
	number = 2,
	year = 1992,
	pages = "712--736"
}

@ARTICLE{minnotte97,
  author = "Michael C. Minnotte",
  title = "Nonparametric testing of the existence of modes",
  journal = "The Annals of Statistics",
  volume = 25,
  number = 4,
  year = 1997,
  pages = "1646--1660"
}

@ARTICLE{mukhopadhyay16,
  author = "Subhadeep Mukhopadhyay",
	title = "Large-Scale Mode Identification and Data-Driven Sciences",
	journal = "Electronic Journal of Statistics",
	volume = 11,
	number = 1,
	year = 2017,
	pages = "215--240",
  note = "arXiv:1509.06428v4"
}

@BOOK{oppenheim89,
  author = "Alan V. Oppenheim and Ronald W. Schafer",
  title = "Discrete-Time Signal Processing",
  publisher = "Prentice Hall",
  year = "1989",
  address = "Englewood Cliffs, New Jersey"
}

@ARTICLE{pyke65,
  author = "Ronald Pyke",
  title = "Spacings",
  journal = "Journal of the Royal Statistical Society, Series B",
  volume = 27,
  number = 3,
  year = 1965,
  pages = "395--449"
}

@ARTICLE{rufibach10,
  author = "K. Rufibach and G. Walther",
  title = "The block criterion for multiscale inference about a density, with
    applications to other multiscale problems",
  journal = "Journal of Computational and Graphical Statistics",
  volume = 19,
  number = 1,
  year = 2010,
  pages = "175--190"
}

@ARTICLE{ruta00,
  author = "Dymitr Ruta and Bogdan Gabrys",
  title = "An overview of classifier fusion methods",
  journal = "Computing and Information Systems",
  volume = 7,
  year = 2000,
  pages = "1--10"
}

@INPROCEEDINGS{siffer18,
  author = "A. Siffer and A. Termier and P. A. Fouque and C. Largouet",
	title = "Are Your Data Gathered? {T}he Folding Test of Unimodality",
	booktitle = "24th ACM SIGKDD International Conference on Knowledge Discovery",
	address = "London",
	year = 2018,
	month = aug,
	pages = "2210-2218",
	note = "doi: 10.1145/3219819.3219994"
}

@ARTICLE{silverman81,
  author = "B. W. Silverman",
  title = "Using kernel density estimates to investigate multimodality",
  journal = "Journal of the Royal Statistical Society, Series B",
  volume = 43,
  number = 1,
  year = 1981,
  pages = "97--99"
}

@ARTICLE{venter67,
  author = "J. H. Venter",
  title = "On Estimation of the Mode",
  journal = "The Annals of Mathematical Statistics",
  volume = 38,
  year = 1967,
  pages = "1446--1455"
}

@ARTICLE{xu92,
  author = "Lei Xu and Adam Kryzy{\.z}ak and Ching Y. Suen",
  title = "Methods of combining multiple classifiers and their applications to
           handwriting recognition",
  journal = "IEEE Transactions on Systems, Man, and Cybernetics",
  volume = 22,
  number = 3,
  month = may,
  year = 1992,
  pages = "418--435"
}

@ARTICLE{kaplansky45,
  author = "Irving Kaplansky and John Riordan",
  title = "Multiple Matching and Runs By The Symbolic Method",
  journal = "The Annals of Mathematical Statistics",
  volume = 16,
  year = 1945,
  pages = "272--277"
}

@ARTICLE{wald40,
  author = "A. Wald and J. Wolfowitz",
  title = "On A Test Whether Two Samples Are From The Same Population",
  journal = "The Annals of Mathematical Statistics",
  volume = 11,
  year = 1940,
  pages = "147--162"
}

@ARTICLE{xu14,
  author = "Ling Xu and Edward J. Bedrick and Timothy Hanson and Carla Restrepo",
	title = "A Comparison of Statistical Tools for Identifying Modality in Body
           Mass Distibutions",
  journal = "Journal of Data Science",
	volume = 12,
	year = 1024,
	pages = "175--196"
}

\begin{appendix}

\section{K Samples}
\label{app:ksample}

Samples K1 through K9 have four modes and K10 through K15 five.  They are
defined in Table~\ref{tbl:kdef} using normal variates N($\mu, \sigma$)
with mean $ \mu $ and standard deviation $ \sigma $, uniform variates
U($a, b$) between the bounds, binomial variates B($N, p$) that have success
probability $ p $ for $ N $ trials, gamma variates G($r, \lambda$) that have
shape $ r $ and rate $ \lambda $, and Weibull variates W($a, b$) with
scale $ a $ and shape $ b $.  The uniform draw is used to provide a fixed
background.  K9 and K14, using binomial variates, have discrete values.

\begin{table}
\centering
\caption{\label{tbl:kdef} K Sample Definitions}
{\small
\begin{tabular}{lllll}
K1  &  60 N($ -1.5 $, 0.25) & 140 N(0, 1)
    & 140 N(4, 1)           &  60 N(5.5, 0.25)      \\
K2  &  50 N($ -5 $, 0.5)    &  50 N($ -2.5 $, 0.75)
    &  50 N(1, 1.25)        &  50 N(4.5, 1)         \\
K3  &  75 N($ -2 $, 1)      &  50 N($ -0.75 $, 0.2) 
    &  50 N(0.75, 0.2)      &  75 N(2, 1)           \\
K4  &  75 N($ -2.5 $, 1)    &  50 N($ -0.75 $, 0.2)
    &  50 N(0.75, 0.2)      &  75 N(2.5, 1)         \\
K5  & 100 N($ -2.5 $, 1.5)  &  50 N($ -0.75 $, 0.3)
    &  50 N(0.5, 0.2)       &  75 N(2, 1)           \\
K6  &  45 N(0, 0.2)         &  30 N(1, 0.15) 
    &  55 N(2, 0.2)         &  40 N(3, 0.25)        \\
    & 130 U($ -1.5 $, 4.5)  \\
K7  &  40 W(2, 0.75)        & 100 W(4, 7) 
    &  50 W(5, 2)           &  60 W(8, 4)           \\
K8  &  60 N($ -2 $, 2)      &  40 W(2, 1) 
    &  40 N(7, 1)           &  60 G(14, 4)          \\
K9  &  60 B(300, 0.04)      &  40 B(300, 0.12) 
    &  60 B(300, 0.20)      &  40 B(300, 0.29)      \\
K10 &  60 N($ -1.5 $, 0.25) & 140 N(0, 1)
    &  50 N(2, 0.25)        & 140 N(4, 1)           \\
    &  60 N(5.5, 0.25)      \\
K11 &  70 N($ -5 $, 0.2)    &  45 N($ -3.5 $, 0.5)
    &  45 N($ -1 $, 0.75)   &  45 N(2, 0.75)        \\
    &  50 N(5, 1)       \\
K12 &  50 N($ -3.5 $, 0.2)  &  50 N($ -2 $, 0.5) 
    &  50 N(0, 0.75)        &  50 N(2.5, 1)         \\
    &  50 N(5, 1.25)    \\
K13 &  55 N(0, 0.2)         &  30 N(1, 0.15)
    &  60 N(2, 0.25)        &  40 N(3, 0.2)         \\
    &  50 N(4, 0.15)        & 115 U($ -1.5 $, 5)    \\
K14 &  60 B(200, 0.05)      &  40 B(200, 0.15) 
    &  60 B(200, 0.25)      &  50 B(200, 0.35)      \\
    &  40 B(200, 0.45)  \\
K15 &  75 N($ -5 $, 1)      &  50 N($ -3 $, 0.2) 
    & 100 N(0, 2)           &  50 N(3, 0.2)         \\
    &  75 N(5, 1)       \\
\end{tabular}
}
\end{table}

\end{appendix}

\end{document}